# On the Use of the Microtremor HVSR for Tracking Velocity Changes:

## A case study in Campo de Dalías basin (SE Spain)

By

Helena Seivane [1] [*], Antonio García-Jerez [1], Manuel Navarro [1]

Luis Molina [2] and Francisco Navarro-Martínez [2]

(1) Department of Chemistry and Physics, University of Almería, 04120 Almería, Spain.
(2) Department of Biology and Geology, University of Almería, 04120 Almería, Spain.

(*) Corresponding author: Helena Seivane Ramos, e-mail: helenaseiv@outlook.com

Abstract

The stability of the low-frequency peaks (< 1 Hz) obtained in the passive seismic survey of Campo de Dalías basin (CDB) by applying the horizontal-to-vertical spectral ratio (HVSR) method was investigated. Three temporary seismic stations were installed in remote sites that enabled studying the stationarity of their characteristic microtremor HVSR (MHVSR) shapes. All stations began to operate in mid-2016 and recorded at least one year of continuous seismic ambient noise data, having up to two years in some. Each seismic station counted with a monitored borehole in their vicinity, registering the groundwater level every 30 minutes. The MHVSR curves were calculated for time windows of 150 s and averaged hourly. Four parameters have been defined to characterize the shape of the MHVSR around the main peak and to compare them with several environmental variables. Correlations between MHVSR characteristics and the groundwater level showed to be the most persistent. The robustness of MHVSR method for applications to seismic engineering was not found to be compromised since the observed variations were within the margins of acceptable deviations. Our results widen the possibilities of the MHVSR method from being a reliable predictor for seismic resonance to also being an autonomous monitoring tool, especially sensitive to the S-wave modifications.

# 1. Introduction

The use of MHVSR in hazard assessments by seismic ambient noise measurements has been exploited to the most extent in a myriad of studies. As stated by Molnar et al. (2018), its standardization has not always been accompanied by a full understanding of the theory behind it. It was mainly the ease in the application what captivated practitioners.

Despite being originally born as a technique for predicting the dynamic response of soils under seismic loading (Nogoshi and Igarashi, 1970-71; Nakamura 1989), the MHVSR evolved to be also a geophysical exploration tool. For such progress to occur, it was necessary defining inverse schemes able to relate theoretically the MHVSR to the soil layering characteristics and their elastodynamic properties (e.g., Arai & Tokimatsu, 2000 2004; Fäh et al., 2001 2003; Lunedei & Albarello, 2010 2015, Sánchez-Sesma et al., 2011).

Reliability of field measurements is often assured by taking into consideration unfavourable circumstances like bad weather, disturbances, and nearby structures (SESAME, 2004). A robust representation of the study site is assured by assessing the reliability of the final MHVSR curve from a statistical point (Chatelain et al., 2008; Cox et al., 2020). Then, a value for the resonance frequency of S-waves may be picked with confidence.

Obtaining reliable estimates of the full shape of the MHVSR curves is key for inversion (see e.g., Kawase et al., 2015; Thabet, 2019 or Perton et al. 2020). Recent studies like those of La Rocca et al. (2020) have pointed to the fact that, while one of the main assumptions here is the stability of the MHVSR, complex geological structures or rough topographies may lead the experimental curves to be far from it. Moreover, results by Lotti et al. (2018) showed time variations on the MHVSR even when the meteorological factors were excluded. For these authors, the subsoil prop-erties of an unstable rockslide were behind the observed changes detectable by the MHVSR shape. Despite observing a clear correlation between some of their MHVSR peak amplitudes and atmospheric temperature, seven months of analysis were insufficient for them to observe some cyclical behaviour.

Seasonal variations on soil properties, specifically on their seismic velocities, are not a new phenomenon for the seismic-engineering community. They have been long studied on array experiments which demonstrated that seismic noise correlations are sensitive to changes in the seismic velocities of the subsurface medium (see e.g., Meier et al. 2010; Hillers et al. 2014-2015, Voisin et al., 2016 or Clements and Donelle, 2018). The mechanism on how these changes are induced is what is holding current discussions. Some authors point to source mechanisms like thermoelastic strains or solid earth tides (Snieder et al., 2002; Takano et al., 2014; Hillers et al. 2014-2015; Kouznetsov et al., 2016; Lotti et al. 2018; Chaput et al., 2018; Miao et al., 2019). Nevertheless, others find in the variability of wavefield composition or source location the explanation for their observations (Kraeva et al., 2009; Stutzman et al., 2009; Zhan et al., 2013; Lepore et al., 2016; Gualtieri et al., 2018; Dybing et al., 2019; La Rocca et al., 2020). Thus, in plain words, studies on long-term monitoring by observation of seismic ambient noise are divided into those tracking the medium behaviour and those which do so for the ambient noise sources variability.

The main motivation of this paper is to investigate the origin of the slight variability in the MHVSR shape found in experimental studies. To this aim, the case of Campo de Dalías basin (CDB), a coastal plain in SE Spain, is analysed. That area has been the subject of numerous structural and hydrogeological investigations since the 70s due to the economic relevance of the intensive agricultural activities. Up to two years of continuous seismic data recorded at three broad-band stations have been processed. Atmospheric, oceanic, and groundwater-table data were also analysed to compare and understand the periodicities observed in the long-term variability of MHVSR shape. The sensitivity of this curve to variations in the elastodynamic properties of the ground structure in CDB has been checked by means of synthetic tests.

2. Study Area and Data Analysis

The precedent case study which motivated research into the MHVSR stability was the one of El Ejido town, where a microzonation based on seismic ambient noise measurements was carried out (García-Jerez et al., 2019). Placed on a sedimentary basin in the Southeast part of the Betic Cordilleras (Figure 1), results of El Ejido city showed long fundamental site periods in good accordance with the sediment thicknesses known to exist there (Pedrera et al., 2015). With an insight of modeling the whole basin as rigorously and trustworthy as possible, the stationarity of MHVSR curves was investigated. Results from urban microzonation showed many MHVSR curves characterised by a broad peak. In this study, the EJDN station (Figure 1) provides a good example of this typology. This kind of MHVSR peak-shapes is commonly not recommended to estimate resonance frequencies. Their stationarity analysis will enable us to better assess the true representative character of these curves in similar geological contexts.

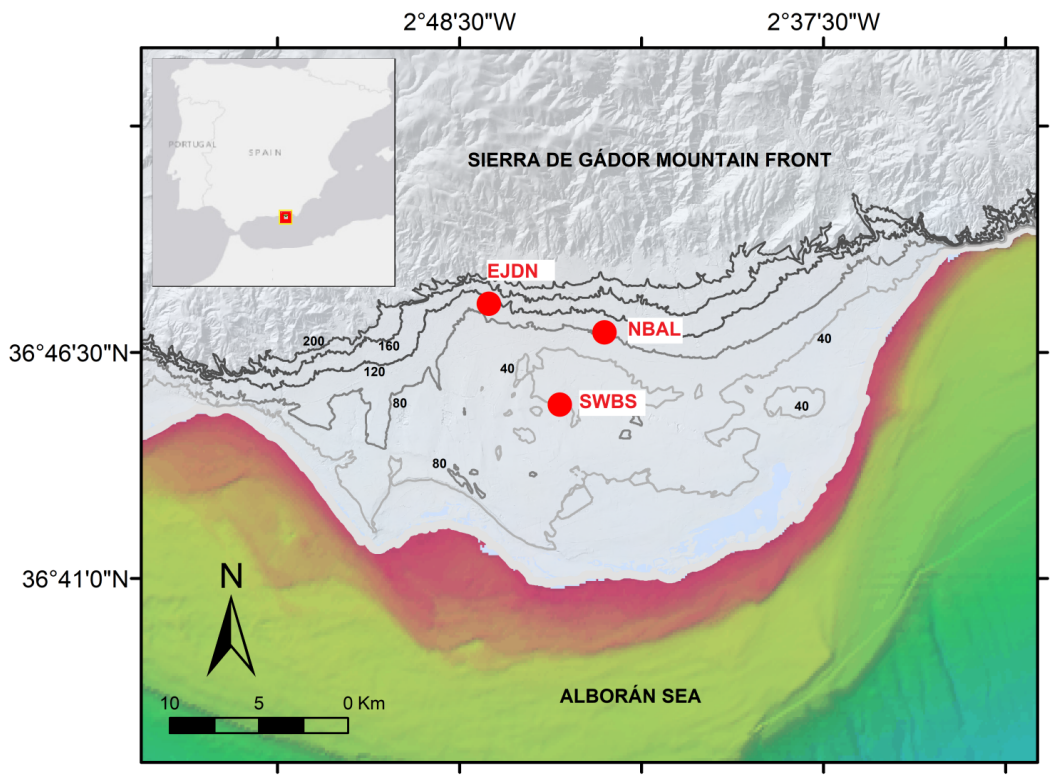

Figure 1. Map of the study area: Campo de Dalías area in Almería (SE Spain). Red circles indicate the position of the broadband seismic stations and piezometers. They were recording respectively seismic ambient noise and underground water level from early 2017 to mid-2019. The line contours represent the topography every 40 metres.

The customary division of seismic noise sources tends to be split into two: natural and human-made sources, where each group is considered to radiate seismic waves in different frequency bands. Despite grouping sources according to their origins being largely accepted (Table 1, Bonnefoy-Claudet et al., 2006), it is certainly true that their frequency boundaries of action are still, and will keep being, fuzzy. The strong site dependence impedes the generalization of frequency boundaries valid everywhere. Only 'soft' statements as "short-period noise is dominated by human activity, while mid- and long-period noise is dominated by effects from ocean waves" by (Nakata et al., 2019) can be made. Thus, it would be expected that in the study area both urban and marine origins predominate over the other more punctual sources.

Both kinds of signals are known to have temporal variations of spectral amplitude and frequency content. In the case of human activity, such patterns are globally observed with day-night and week-weekend cycles that match with the specific rhythm and lifestyle of the areas under study. The recent global lockdowns due to the COVID-19 pandemic have demonstrated that even reference low noise laboratories like the Black Forest Observatory in Germany showed noise level reductions on the frequency band between 4 to 14 Hz. The exceptional 2020 quiet period came to demonstrate how anthropogenic sources get to affect large areas (Lecocq et al., 2020).

As for the action of the near Alborán Sea, in the western Mediterranean, this would be expected to be more intense in the band of long periods (3 to 300 s, Ardhuin et al., 2015) generated by the oceanic microseisms (Longuet-Higgins, 1950; Nishida, 2017). Despite being this microseism range far from the usual of engineering interest (Albarello and Lunedei, 2011), the resulting MHVSR curves in El Ejido town with peaks as low as 0.4 Hz lead to think on their possible effect on the MHVSR shapes. Two years of single-station measurements were analysed with the purpose of clearing up the doubtful action of such sources over the MHVSR shapes with clear peaks well below 1 Hz. The stations were operating from mid-2016 to mid-2019 and were installed in sites a few kilometres far from the main urban areas.

Secondary microseisms (SM) acting in the band from 0.1 to 0.4 Hz were proved to have stronger stability in time and to be more influenced by the local climate than by global patterns as the primary microseisms (PM) between 0.03 and 0.1 Hz demonstrably do (Stutztman et al., 2009; Carvalho et al., 2019). The local climatic variables of air temperature, barometric pressure, and wind speed were obtained from the series repository of the Spanish Agency of Meteorology ([www.aemet.es](www.aemet.es), AEMET). Concerning sea level, it was obtained by the harmonic constituents at the tide gauge of Almería port given by the national Spanish State Port Agency ([www.puertos.es](www.puertos.es), Puertos del Estado). Figure 2 shows their raw data set together with the raw piezometric data from SWBS station. The correlation analysis to assess the relationship between these climatic variables and the MHVSR parameters was done by using the built-in MATLAB function *corrcoef*. The filtering of raw data series, both MHVSR and climatic variables, was performed by median and Butterworth filters implemented in the built-in MATLAB functions *movmedian* and *butter* respectively.

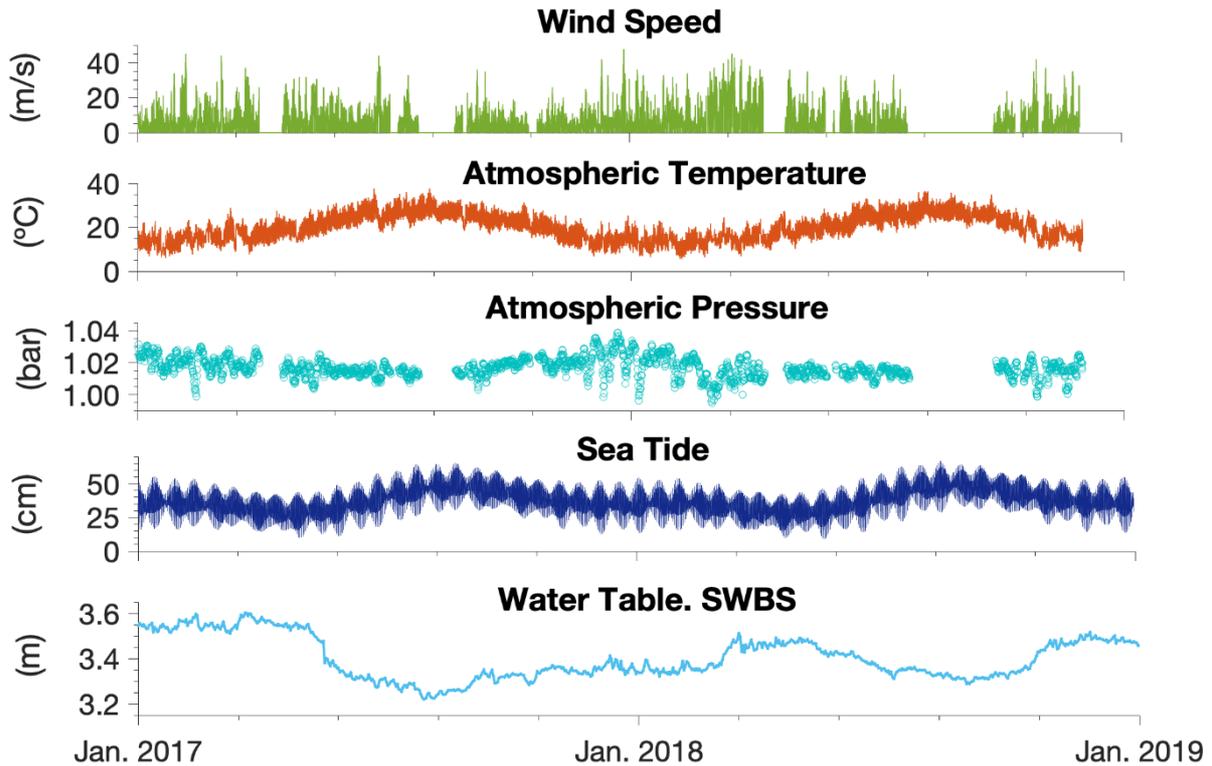

Figure 2. Data series of weather variables, sea tide, and groundwater levels assessed in this study in the period between 2017 and 2019.

Known to be a semi-arid area, Campo de Dalías has an overexploited system of aquifers due to the intensive agriculture activities taking place there (Pulido-Bosh et al., 2020). Piezometric observations were acquired at boreholes located a few metres away from the long-term seismic stations (EJDN, NBAL, and SWBS stations in Figure 1) to further investigate possible implications for MHVSR stability. The sensors utilised for these long-term observations of seismic ambient noise were broadband seismometers Güralp 3ESPDC. The technical specifications of these broadband sensors assure levels of self-noise well under the NLNM (New Low Noise Model, Peterson, 1993) for all their components on our frequency band of interest (0.1 to 1 Hz). None of the instruments were installed on the bare ground. These three stations, with a long-term outlook, were vault mounted and sheltered. The station EJDN was the only one buried a half metre deep, and therefore it would be considered as our highest quality data acquisition according to the criteria by Foti et al. (2018).

All the MHVSR analysis done over the seismic ambient noise records was performed by means of MATLAB programming. The seismic ambient noise signals were baseline corrected and windowed with a length of 150 s plus an overlap of 50%. Each time window was energy-normalized according to the criterion proposed by Sánchez-Sesma et al. (2011) so as to follow more closely the diffuse approach under which the MHVSR inversions were eventually done. A Hamming window was applied in the Short-time Fourier transform (STFT) to obtain the frequency spectrum of each of the three motion components. Smoothing of these Fourier spectra was done through the function proposed by Konno and Ohmachi (1998) with a bandwidth value of 40. The average of the horizontal spectrum was defined by the vector summation of the two individual components (Albarello and Lunedei, 2013).

The synthetic approach to assess the impact of seismic velocity changes on the MHVSR shapes was performed by the freeware software HV-Inv (García-Jerez et al., 2016). Synthetic tests were carried out by using the layered earth models obtained from inversion of the mean MHVSR curves at each station. The ranges of thicknesses and elastic parameters for each layer were previously bounded by all the geological and geophysical information gathered for the study of the basin. Being surrounded by so many wells, with available stratigraphical descriptions, also helped to get simple but representative 1-D earth models that reproduced the experimental MHVSR curves.

Assessment of the MHVSR peak-shape variability was performed by observing changes in amplitude, frequencies, and width (Figure 3). The former properties, amplitude, and frequency are defined as those where peaks reach their maximum value. The tracking of troughs associated with each MHVSR peak was also incorporated in the stationarity analysis. They are defined as the first minimum after the MHVSR peak. As proposed by Piña-Flores et al. (2020), MHVSR troughs may be considered as new parameters to help in phase velocity analysis from multistation experiments. Therefore, they turn to be also related to velocity models.

Definition of width as a property, knowing how different may look MHVSR typologies, required adaptation. That is, a reference amplitude value for each peak analysed was pre-defined (Figure 3). Given this reference value, the width is defined as the frequency band covered between the two points crossing the MHVSR curve at the reference amplitude. The reference amplitudes chosen in this study for assessment of each MHVSR peak are shown in Figure 4. They correspond to MHVSR amplitudes of 2.05, 3.0, and 3.5 for EJDN, NBAL, and SWBS stations respectively. The selection of these amplitudes followed criteria based on the stability in the iterative searching and data analysis algorithms. Since quantification of velocity changes is not aimed in this work, but rather observing MHVSR shape changes, no general rule for selecting reference amplitudes is here defined.

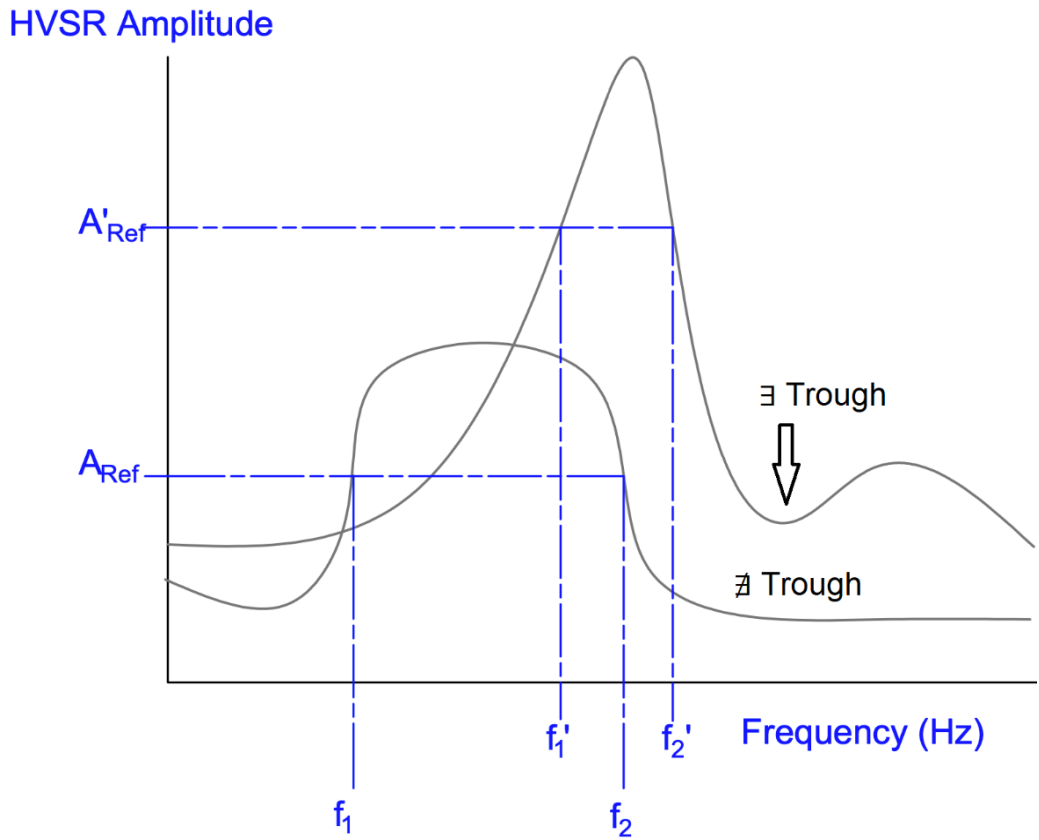

Figure 3. Scheme for definition of MHVSR peak-width and trough. Given a MHVSR peak the width is linked to an amplitude of reference A'$_{Ref}$ and A$_{Ref}$ in these examples. The widths from these two MHVSRMHVSR curves are defined then as $f_2-f_1$ and $f_2' - f_1'$ respectively. The definition of troughs is also exemplified; while one curve has a minimum after its main MHVSR peak, the other lacks from it.

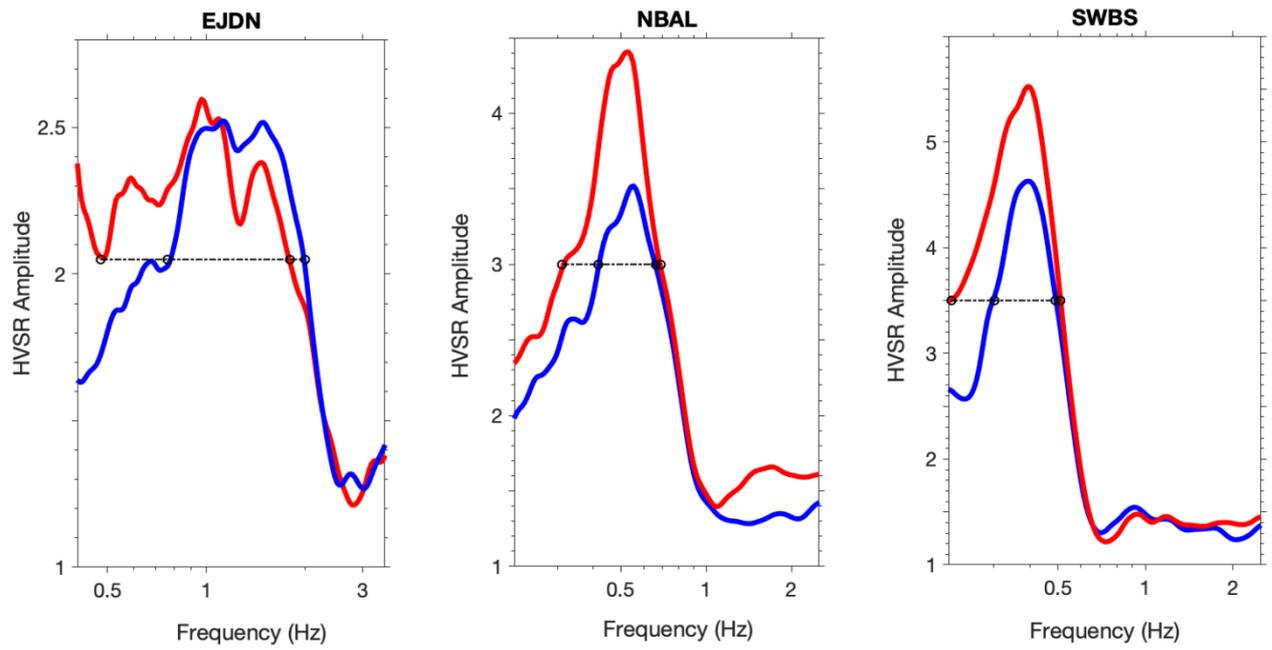

Figure 4. Example of the reference amplitudes chosen in this study. They are marked with black lines over each MHVSR curve. Red curves correspond to the 10-hours averaged MHVSR curve for the time window with the widest peaks. The blue ones correspond to the 10-hours averaged MHVSR curve for the time window with the narrowest peaks found at each station.

## 3. Results and Discussion

### 3.1 Synthetic Variations

The dependence of the MHVSR shape on the ground model parameters is addressed in this section. The 1-D elastodynamic models, over which modifications were introduced, are presented in Table 1. These models resulted from inversion of the average MHVSR curve at each station. The parameterisation strategy was in line with the recent guidelines of Foti et al. (2018). Previous knowledge of mean thicknesses, densities, and body-wave velocities for the main sedimentary units found in this area (Marín-Lechado et al., 2005; Marín-Lechado, 2005) was key. It helped in defining the boundaries of the parameter space. Figure 5 summarises the degree of agreement between the observed and the modeled MHVSR curves.

Table 1. 1-D models obtained from the MHVSR inversion and previous geological and geophysical constraints for the three study sites: SWBS, NBAL, and EJDN (locations in Figure 1).

|     | Thickness (m) | $V_P$ (m/s) | $V_S$ (m/s) | $\rho$ (kg/m$^3$) | Poisson |
|-----|---------------|-------------|-------------|-------------------|---------|
| **SWBS Station** | | | | | |
| (1) | 10            | 1278        | 597         | 2000              | 0.36    |
| (2) | 197           | 1240        | 661         | 2350              | 0.30    |
| (3) | 463           | 2143        | 1029        | 2250              | 0.35    |
| (4) | 55            | 3230        | 1552        | 2400              | 0.35    |
| (5) | -             | 5961        | 2531        | 2700              | 0.39    |
| **NBAL Station** | | | | | |
| (1) | 10            | 819         | 432         | 2000              | 0.31    |
| (2) | 196           | 1496        | 719         | 2350              | 0.35    |
| (3) | 365           | 2496        | 1200        | 2250              | 0.35    |
| (4) | 60            | 3198        | 1668        | 2400              | 0.31    |
| (5) | -             | 4412        | 2401        | 2700              | 0.29    |
| **EJDN Station** | | | | | |
| (1) | 10            | 838         | 500         | 2000              | 0.22    |
| (2) | 115           | 1329        | 802         | 2250              | 0.21    |
| (3) | 120           | 1962        | 1097        | 2400              | 0.27    |
| (4) | 578           | 3030        | 1771        | 2400              | 0.24    |
| (5) | -             | 5100        | 2466        | 2700              | 0.28    |

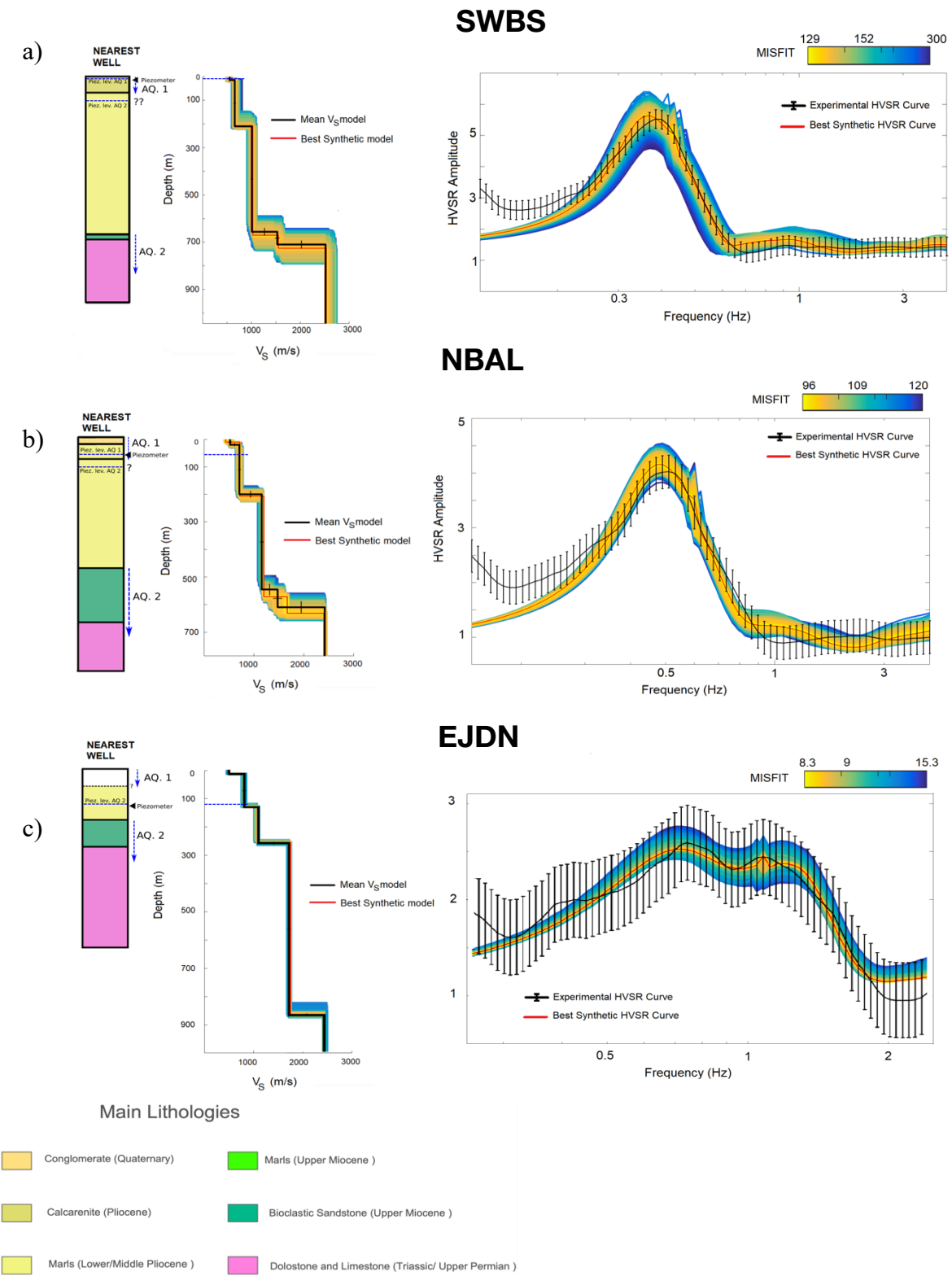

Figure 5. Inversion of the experimental MHVSR curves that are representative of sites a) SWBS b) NBAL and c) EJDN. The theoretical MHVSR curves and their $V_S$ models tested during the inversion process are coloured according to the misfit bars. Best models achieved are highlighted in red (Table 1). The stratigraphic columns from the nearest deep boreholes of each study location are shown in the left panels. The depth of the loggers used for piezometric observations is marked with a black triangle. Average piezometric levels are shown with dashed blue lines. Blue arrows show the layers that make up the AQ1 and AQ2 aquifers.

The model from the station EJDN needs a special comment. Despite having been discussed and hypothesised to be a likely consequence of complex wavefields near the edges of basement slopes, strong lateral thickness variations, fractured bedrock conditions or 2-D resonances (see e.g., Guillier et al., 2007; Bonnefoy-Claudet et al., 2008; Le Roux et al., 2012; Galgaro et al., 2014; Gosar, 2017 or Okamoto & Tunno, 2018; Sgattoni & Castellaro, 2020), broad peaks may also be reproduced by 1-D models (Figure 5.c). Analysis of directivity in the MHVSR shows approximate azimuthal independence below 3 Hz, which points to the irrelevance of 2D-3D effects caused by the deep structure. All boreholes in the northern limit of Campo de Dalías, where EJDN is located, have the metamorphic bedrock at depths no greater than 300 metres. The relatively low seismic velocity of the model at this depth would be an indicator of the grade of fracturing or karstification of these units, but that discussion is far from the objective of this study. The five-layer structure associated with EJDN station complies with our prime aim of studying the effect of velocity changes on the MHVSR shapes.

For the sake of simplicity, only adjacent layers had their seismic velocities simultaneously modified. As well, two strategies were followed when modifying the velocities of each layer. The first strategy consisted in letting the Poisson coefficient constant while the seismic velocities were modified, thus modifying both $V_S$ and $V_P$ in the same proportion. The second approach involved varying the Poisson coefficient in order to comply with a more realistic situation, where both wave velocities may have a different rate of variability (e.g., O'Connell R., 1974). In no case were the densities or thicknesses modified.

Figure 6 presents the variations experienced by the MHVSR curves after introducing in each layer, one by one, increments and decrements on the seismic velocities following the former strategy. The rates of introduced change in the seismic wave velocities of the reference models (Table 1) varied between -10% and +10 %. Only when changes greater than 10 % were measured in the MHVSR parameters of interest (Amplitude, frequency, width, and trough), were then deemed as actual variations. Table 2 gathers all the results measured under such criterion for variations introduced in single and multiple layers. As highlighted in Figure 3, the existence of an MHVSR trough cannot be always assured. The synthetic MHVSR linked to NBAL station does not comply with the criteria of having a clear minimum point right after its maximum. Therefore, no measure was done on the behaviour of this parameter neither in the synthetic nor in the experimental case of NBAL.

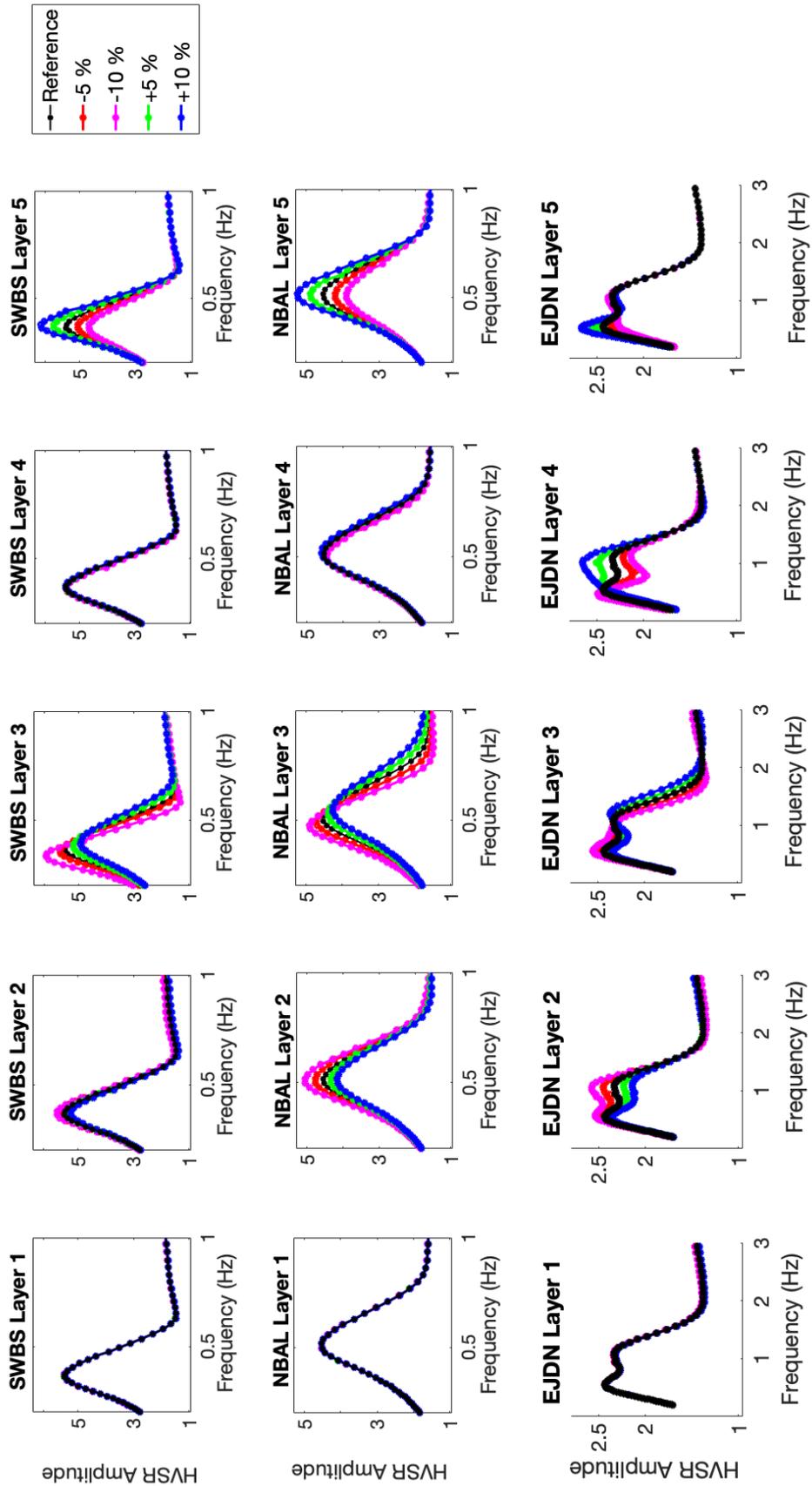

Figure 6. Synthetic MHVSR variations after modifying one single layer following the procedure of maintaining a constant Poisson coefficient while varying the $V_S$ velocities.

Table 2. Summary of results observed on the synthetic MHVSR curves of each study site under variations of the seismic velocities between -10% and +10% from the reference values (Table1). These variations followed the procedure of maintaining a constant Poisson coefficient. Each row identifies the variation of a single ground layer or combination of consecutive layers. The minus symbols - correspond to an observed variation of less than 10% of the reference MHVSR parameter. The plus + symbols mark MHVSR variations higher than 10%.

| Layer Combination | Amplitude | | | Width | | | Frequency | | | Trough | |
|---|---|---|---|---|---|---|---|---|---|---|---|
| | SWBS | NBAL | EJDN | SWBS | NBAL | EJDN | SWBS | NBAL | EJDN | SWBS | EJDN |
| (1) | - | - | - | - | - | - | - | - | - | - | - |
| (2) | - | + | - | - | + | - | + | - | + | - | - |
| (3) | + | + | - | - | + | + | + | + | + | + | + |
| (4) | - | - | - | - | + | - | - | - | + | - | - |
| (5) | + | + | + | + | + | - | - | - | + | - | - |
| (1) & (2) | - | + | - | - | + | + | - | - | + | - | - |
| (2) & (3) | + | + | + | + | + | + | + | + | + | + | + |
| (3) & (4) | + | - | - | + | - | + | + | + | + | + | + |
| (4) & (5) | + | + | + | + | + | + | - | - | + | - | - |
| (1) & (2) & (3) | + | + | + | + | + | + | + | + | + | + | + |
| (2) & (3) & (4) | + | + | + | + | + | + | + | + | + | + | + |
| (3) & (4) & (5) | + | + | - | + | + | + | + | + | + | + | + |
| (1) & (2) & (3) & (4) | + | + | + | + | + | + | + | + | + | + | + |
| (2) & (3) & (4) & (5) | - | - | - | + | + | + | + | + | + | + | + |
| ALL | - | - | - | + | + | + | + | + | + | + | + |

At a first glance, it is easy to note how by modifying only the first layer (Labelled (1) in Table 1) none of the four MHVSR parameters are affected. Only the high-frequency features depend strongly on this layer. The behaviour of the three MHVSR peaks is also coincident under simultaneous modification of the upper four layers and all layers.

The link found between MHVSR-shape parameters and ground model properties can provide some insight into the location of possible model alterations, causing the experimental variations found in the MHVSR in later sections. The low sensitivity of the MHVSR, in the frequency range shown, to variations in some layer parameters discards the tracking of the main peak features as a fully autonomous technique to quantify velocity changes of the medium. Support by other direct or indirect methods would be recommended for this purpose.

Figure 7 is analogous to Fig. 6 but shows the results for the second strategy. The $V_S$ values were kept fixed and the Poisson coefficients were varied between -50 % and 20% with respect to the reference values (Table 1). Compared to what is observed in Fig. 6, layers (2) and (3) are still the ones with a higher influence on the MHVSR peak shape. But contrarily to $V_S$ changes, the variation in $V_P$ values of the deepest layers does not make any alteration. The variations observed in Fig. 7 have to do mainly with the MHVSR amplitude. The peak and trough frequencies seem to be only altered by $V_S$ modifications (see Fig. 6). This is shown in Table 3, where it can be seen that for NBAL and SWBS stations the frequency-peak variations are detected in none of the studied combinations. The same cannot be concluded for EJDN station. The strong amplitude variations make the MHVSR peak frequency oscillate between the two local maxima comprised within the broad-peak frequency range. Figure 8 exemplifies such a situation, supporting so the warnings about not considering broad MHVSR peaks as reliable estimators of the resonance frequency of a site.

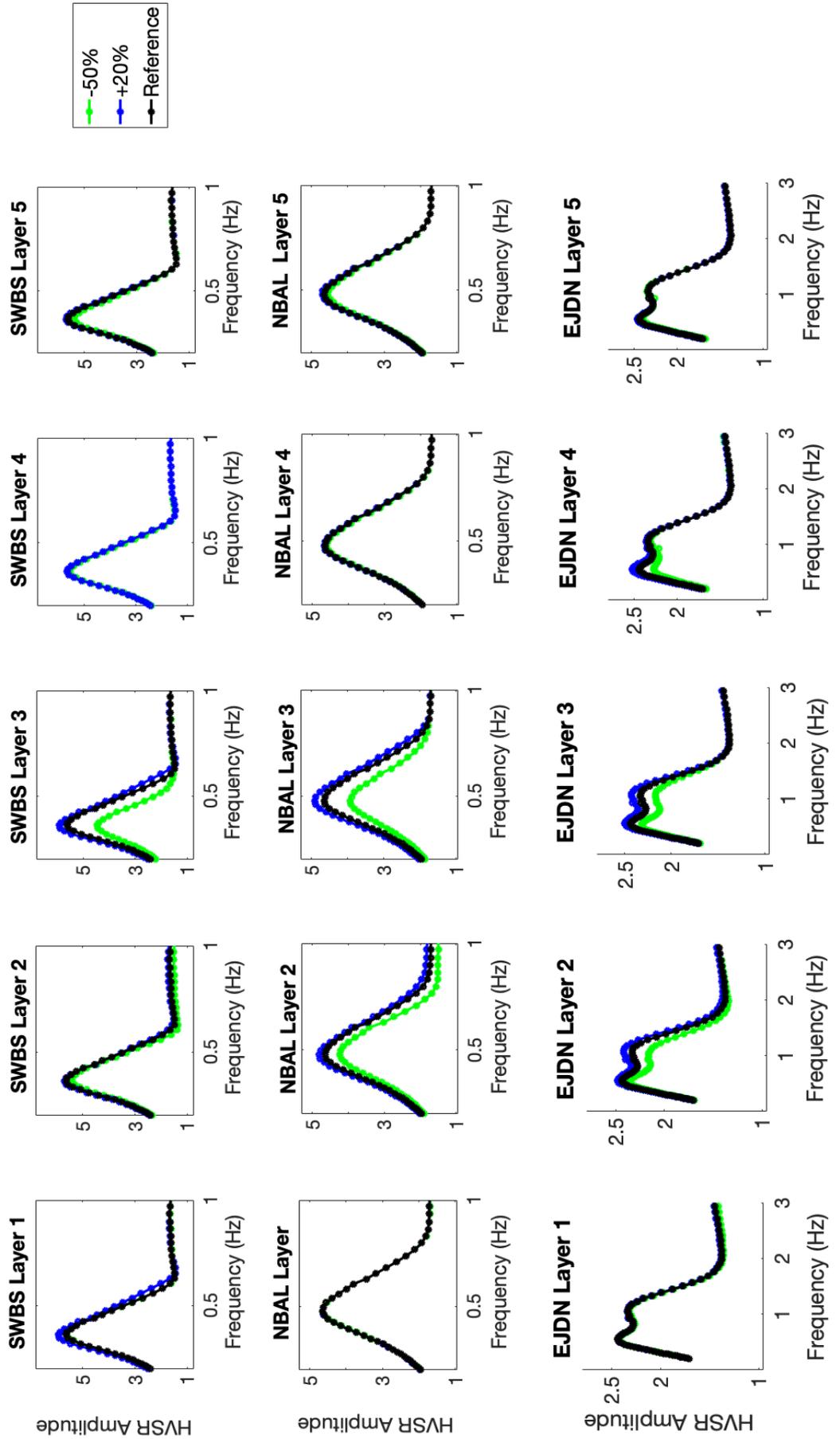

Figure 7. Synthetic MHVSR variations after modifying one single layer following the procedure of varying the Poisson coefficient while maintaining the $V_S$ values unaltered.

Table 3. Summary of results observed on the synthetic MHVSR curves of each study site under variations of the $V_P$ velocities between -50% and 20% from the reference values (Table 1). These variations followed the criteria of variable Poisson. Each row identifies the variation of a single soil layer or combination of consecutive layers. The minus symbols - correspond to observed variation of the reference MHVSR parameter of less than 10% for both -50% and +20% $V_P$ increments. The plus + symbols mark MHVSR variations greater than 10% for any of those $V_P$ increments.

| Layer Combinations | Amplitude | | | Width | | | Frequency | | | Trough | |
|---|---|---|---|---|---|---|---|---|---|---|---|
| | SWBS | NBAL | EJDN | SWBS | NBAL | EJDN | SWBS | NBAL | EJDN | SWBS | EJDN |
| (1) | - | - | - | - | - | - | - | - | - | - | - |
| (2) | - | + | - | - | + | + | - | - | - | - | - |
| (3) | + | + | - | + | + | - | - | - | - | - | - |
| (4) | - | - | - | - | - | - | - | - | + | - | - |
| (5) | - | - | - | - | - | - | - | - | - | - | - |
| (1) & (2) | - | + | - | + | + | + | - | - | - | - | - |
| (2) & (3) | + | + | + | + | + | + | - | - | - | - | - |
| (3) & (4) | + | + | + | + | + | + | - | - | - | - | - |
| (4) & (5) | - | - | - | - | + | - | - | - | + | - | - |
| (1) & (2) & (3) | + | + | + | + | + | + | - | - | - | - | - |
| (2) & (3) & (4) | + | + | + | + | + | + | - | - | - | | - |
| (3) & (4) & (5) | + | + | + | + | + | + | - | - | + | - | - |
| (1) & (2) & (3) & (4) | + | + | + | + | + | + | - | - | - | - | - |
| (2) & (3) & (4) & (5) | + | + | + | + | + | + | - | - | - | - | - |
| ALL | + | + | + | + | + | + | - | - | - | - | - |

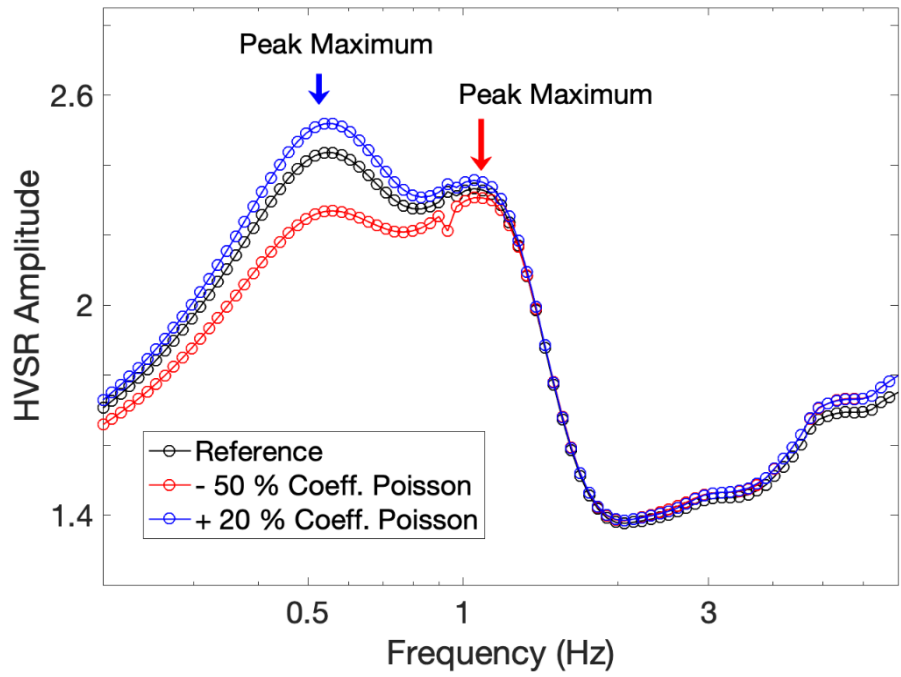

Figure 8. Synthetic MHVSR variations for EJDN model (Table 1). The amplitude variations from its lowest peak part, around 0.5 Hz, make the HVSR peak-frequency oscillate between 0.5 and 1.1 Hz in these synthetic trials.

## 3.2 MHVSR Shape-Variations Observed in Campo de Dalías

The long-term variations of the MHVSR peak parameters were analysed for the stations NBAL, EJDN, and SWBS. Figure 9 contains the MHVSR peak amplitude in the period when each station was continuously operating. Each of the MHVSR peak parameters was sampled hourly for the whole period investigated. The "raw" amplitude series were previously normalised by their mean value and they are shown in black colour in Figure 9.

The bottom panels for each seismic station in Figure 9 are a zoom-in of the top panels. This zoom is aimed to show the normalised peak-amplitude series after applying a sliding median filter with a window length of 30 days. After this filtering, it is easier to differentiate long-term changes. NBAL station is the one that has a clear periodicity since the acquisition in this station enabled us to see up to two years.

The evolution of the four MHVSR parameters, also tracked by synthetic modifications in the previous section, is displayed in Figure 10. NBAL is the station that exhibits the strongest variations, especially seen on the peak-width parameter. The mean values of each MHVSR parameter with their standard deviations for each station are listed in Table 4. But before going into a discussion about the similarities with the synthetic results of the previous section, we will first discuss the correlation with the time series of climate variables.

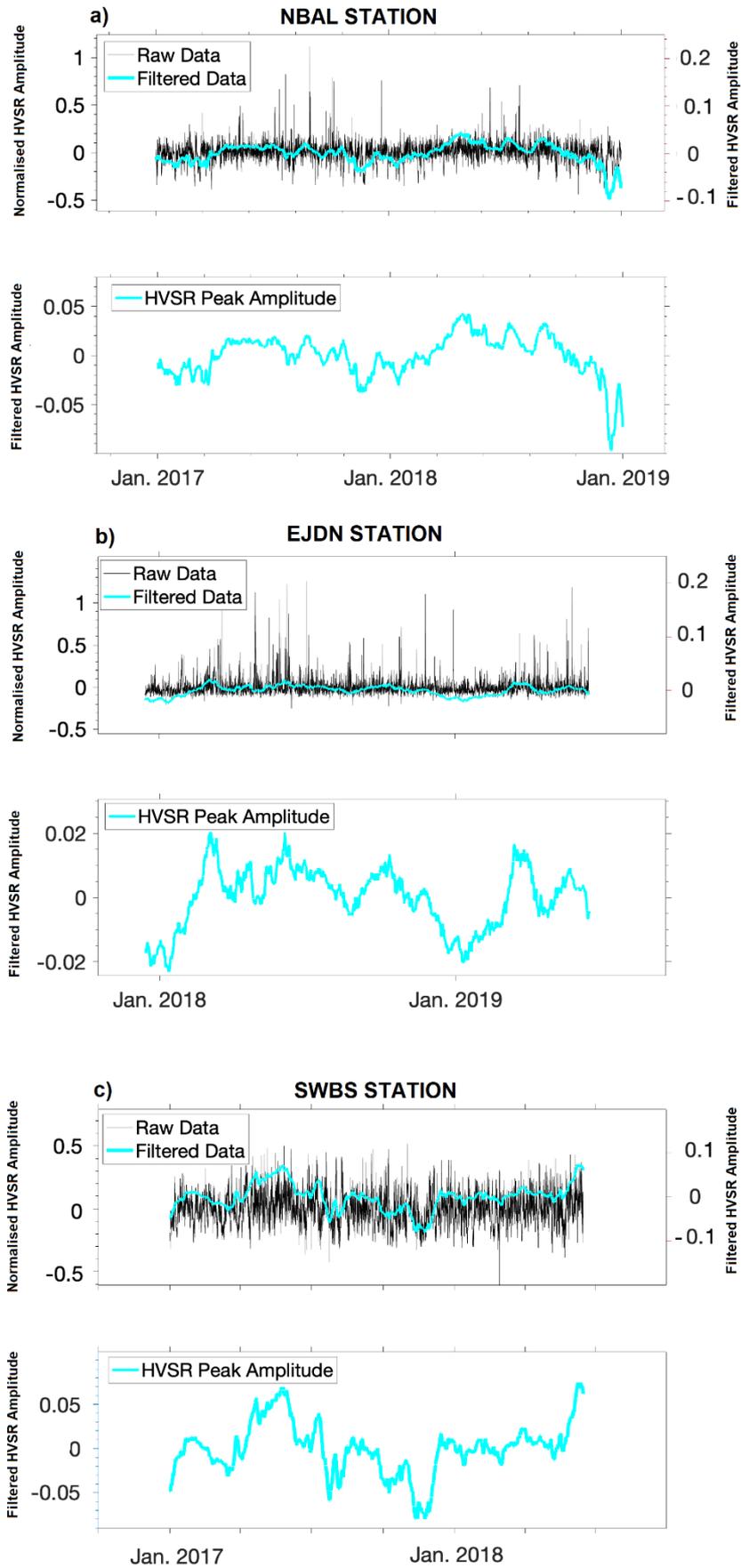

Figure 9. Time series of the MHVSR peak-amplitudes normalised to their mean values for a) NBAL station from early 2017 to mid-2019 b) EJDN station from early 2018 to mid-2019 and c) SWBS station from early 2017 to mid-2018. The cyan curves represent the MHVSR amplitude series after filtering the raw series with a median sliding window 30 days long.

Table 4. Mean values of the MHVSR parameters and their standard deviation during the observation periods shown in Figures 9 and 10.

|      | Amplitude | Width (Hz) | Peak (Hz) | Trough (Hz) |
|------|-----------|------------|-----------|-------------|
| NBAL | 4.7 ± 0.5 | 0.38 ± 0.06 | 0.50 ± 0.05 | - |
| EJDN | 2.7 ± 0.3 | 1.2 ± 0.1 | 1.1 ± 0.2 | 2.7 ± 0.1 |
| SWBS | 5.6 ± 0.7 | 0.24 ± 0.03 | 0.39 ± 0.02 | 0.71 ±0.02 |

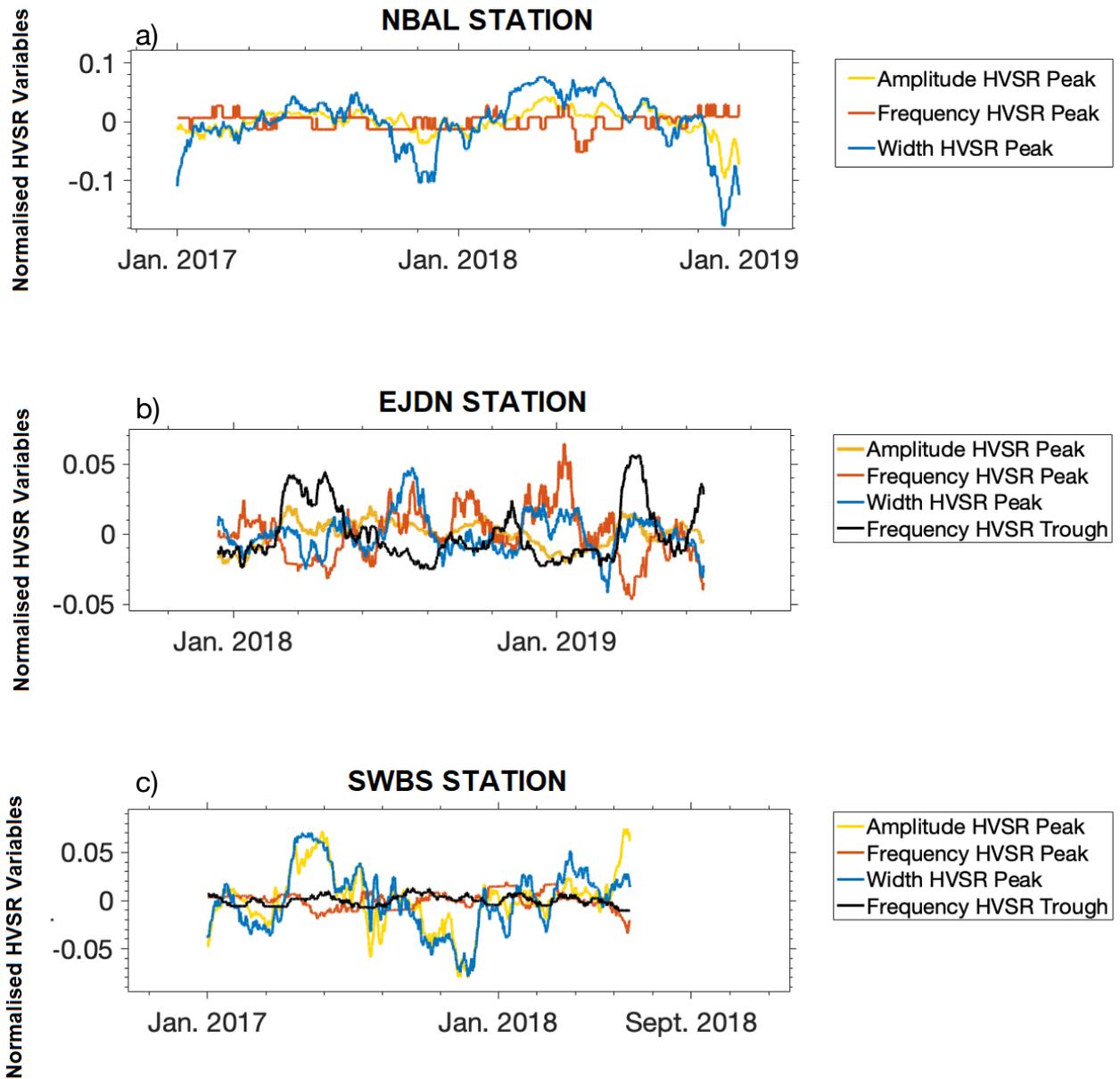

Figure 10. Experimental variations from all MHVSR parameters investigated for a) NBAL station b) EJDN station and c) SWBS station. These series were obtained after applying a moving median filter with a window size of 30 days and normalising to their mean values.

3.2.1 Correlation with weather, oceanic, and groundwater cycles

The quantification of relationships between climate and MHVSR variables was done by calculating the Pearson linear correlation coefficient (abbreviated as CC-P). The hypothesis of linear relationship between the climate variables and the measured changes in MHVSR peak-shapes was so tested. The rejection of the null hypothesis was set at a p-value < 0.05, that is a conditional probability lesser than 5% of rejecting it. Table 5 gathers all the coefficients that resulted for each of the three stations studied during two time intervals. In addition to Pearson correlation, the Spearman correlation was also used to test whether a less restrictive relationship than the linear one could exist. However, the Spearman rank coefficients did not exhibit much difference with the CC-P values shown in Table 5. For the sake of simplicity, since both statistics gave very similar results, Table 5 only shows the CC-P values.

The availability of climate and groundwater data restricted the time length to observe their correlation with the MHVSR curves. Only time periods with continuity of all variables assured were tested. This was necessary to properly compare the grade of influence over the MHVSR between the investigated variables. In some cases, as EJDN case, the climate variables have gaps that impede reliable comparisons for time intervals longer than five months. The longest period with continuous piezometric data at EJDN station lasted from August 2018 to January 2019. Thus, this was the time-scale tested in first place for this station in relation to the water level and other climate variables.

Under the lack of individual meteorological observations at these seismic stations, all of them are assumed to be equally affected by the regional climate conditions in the basin. The gusts of wind recorded by the nearest buoys and meteorological stations, run by Puertos del Estado and AEMET, are taken as the best approximation of what may be affecting each of the three individual locations.

The same happens with temperature and pressure variables. Hence, the water level would be our best approximation to seek for local differences, if existed, between them. Paying attention to the CC-P values in Table 5 some significant relations are brought to light.

The direct interference of wind flow with the seismic sensors has been demonstrated to cause exaggerated spectral ratios at low frequencies through the higher empowering of the horizontal components (Mucciarelli et al., 2005; Barajas-Olalde & Jeffreys, 2014). In fact, the guidelines of SESAME (2004) put a limit of 5 m/s on maximum local wind speed for reliable measurements. They especially emphasize the consequences on the identification of low-frequency resonances (< 1 Hz). On the other hand, recent investigations by Johnson et al. (2019a-b) found that wind effects are also notably present on the band of high frequencies (5 - 200 Hz). According to these authors, wind gusts above 2 m/s can trigger ground motions with PGV values greater than those expected for local earthquakes of magnitudes between 1.0 and 1.5. According to Johnson et al. (2019b), the signal in this high-frequency band can be polluted by the coupling of atmospheric processes through any object on the surface.

However, the wind is the atmospheric variable showing the lowest correlation coefficients with the MHVSR peak parameters for the three stations and the two time periods analysed in this study (Table 5). This is quite remarkable in a wind-prone area like Campo de Dalías, whose mean wind speed for the period 2018-2019 is 5.5 m/s. Our findings do not contradict the other observations previously published on the matter. It should be noticed that these are long-term relationships. Interaction of punctual and strong wind-gusts will clearly blur the MHVSR shapes, especially in poorly isolated stations. In fact, the values presented in Table 5 for the wind speed are brought from moving-median filtering with window length of 3 days. Using a 30-days length for filtering of the wind speed data, as in the rest of the variables, led to accepting the null hypothesis for the relationships between this variable and all the MHVSR parameters.

Table 5. Pearson correlation coefficients (CC-P) between climate variables and MHVSR peak parameters during periods (1) and (2). All data series in this analysis, except for wind speed, were filtered with a median average window of 30 days. The comparison between wind speed series and MHVSR variables was done by using a sliding median window of 3 days.

| | Amplitude | Width | Peak Freq. | Trough Freq. |
|---|---|---|---|---|
| **Wind Speed** | | | | |
| NBAL [1] | 0.282** | 0.236** | -0.064** | - |
| NBAL [2] | 0.288** | 0.205** | 0.130** | - |
| SWBS [1] | 0.338** | 0.354** | -0.248** | -0.139** |
| EJDN [2] | -0.088** | 0.105** | 0.056** | 0.369** |
| **Sea Tide** | | | | |
| NBAL [1] | -0.888** | -0.564** | -0.117** | - |
| NBAL [2] | 0.087** | -0.249** | -0.254** | - |
| SWBS [1] | -0.869** | -0.875** | 0.387** | -0.730** |
| EJDN [2] | -0.263** | 0.127** | 0.419** | -0.307** |
| **Atmospheric Pressure** | | | | |
| NBAL [1] | -0.374** | -0.202** | -0.060** | - |
| NBAL [2] | 0.041$^X$ | -0.278** | 0.349** | - |
| SWBS [1] | -0.667** | -0.700** | 0.132** | -0.502** |
| EJDN [2] | 0.147** | -0.011$^X$ | 0.202** | -0.250** |
| **Atmospheric Temperature** | | | | |
| NBAL [1] | -0.327** | 0.233** | 0.541** | - |
| NBAL [2] | 0.315** | -0.383** | -0.057** | - |
| SWBS [1] | -0.292** | -0.338** | -0.107* | 0.078** |
| EJDN [2] | 0.444** | 0.408** | 0.838** | -0.261** |
| **Groundwater Level** | | | | |
| NBAL [1] | 0.848** | 0.618** | 0.169** | - |
| NBAL [2] | 0.073** | 0.494** | 0.079** | - |
| SWBS [1] | 0.197** | 0.442** | 0.509** | -0.141** |
| EJDN [2] | 0.450** | -0.540** | -0.557** | 0.662** |

[1] Period from March 2017 to August 2017.

[2] Period from August 2018 to January 2019 for EJDN, from September 2018 to January 2019 for NBAL.

\* Corresponds to a level of significance with p-value < 0.05
\*\* Corresponds to a level of significance with p-value < 0.01
$^X$ Corresponds to a level of significance with p-value > 0.05

Contrary to what was expected by Lotti et al. (2018) in their small-scale experiment with seismic stations separated hundreds of metres, our medium-scale area of investigation allows formulating different hypotheses. Experiencing a non-identical degree of influence under the same meteorological variations may be justified by their non-identical soil conditions (Table 1).

The other three climate variables yet not discussed showed correlation coefficients that can be as high as 0.89, like in the case of the sea tide and the peak amplitude at NBAL (Table 5). Despite these high CC-P values found during time windows of several months, it does not seem to be a steady and dominant relationship with any particular variable but complex dependences. When looking at the two periods investigated for NBAL station, it is seen how the CC-P values for tide and pressure (compared with both MHVSR peak frequency and amplitude) change even in sign. In addition, the correlation coefficient between the peak width and the atmospheric pressure also changes its sign from a period to another.

The oceanic interactions causing the well-known microseisms are not either a simple process. In effect, their triggering mechanisms are still under debate. It has been demonstrated that seismic noise levels depend, from sea to sea, on the ocean-site effect and its efficiency to transmit the seismic energy from the ocean to the continent (Beucler et al., 2015; Gualtieri et al., 2015). Thus, to properly assess the microseism print of the Alborán shelf as commonly done (Bromirski et al., 2013), it would be needed a network of ocean bottom seismometers. Counting on both simultaneously, offshore and land-based networks, would be an adequate procedure to discuss the oceanic microseism influence on the coastal seismic stations on CDB.

Under the lack of an offshore network, only light assumptions can be done. But they may anyway be a benefit for preliminary discussion relating SM with the long-term observations on MHVSR curves. SM was proved to propagate mainly by Rayleigh waves in their fundamental mode and be the strongest component of the power spectral density in their frequency band of action. Furthermore, their spectral amplitudes have seasonal variations (Nishida, 2017). But, conversely, to what may be reasoned for the other climate variables, the SM sources are expected to affect equally all seismic stations in Campo de Dalías. SM is not conceived as a mechanism changing the medium properties, but a direct source of seismic waves coming into de medium.

Tidal modulations of the SM have been demonstrated to exist and to be synchronised with the cyclic patterns of energy in short-term periods (Young et al., 2013; Beucler et al., 2015, Becker et al., 2020). Tidal has been proposed as a proxy for SM energy, being the high tides associated with the periods of SM energy increasement. But along with tidal modulations, the short period band of SM ($f > 0.2$ Hz) is known to hold strong correlations with local winds (Hillers and Ben Zion, 2011; Bromirski et al., 2005). Figure 11 shows the energy variation (in arbitrary units) of the seismic ambient noise in the frequency band between 0.3 and 1 Hz for the three stations studied. This total energy estimation was calculated adding up the power spectral densities of the three components across the mentioned frequency band. The characteristic pattern observed holds correlation coefficients larger than 0.67 with the regional wind speed for the three stations (black curve in Figure 11). Comparing the total energy estimates with the MHVSR variations (Figure 10), none of the MHVSR peak parameters got to hold CC-P values larger than those obtained when comparing them to the wind speed (Table 5). Even with a strong correlation between the seismic energy variation and wind speed in CDB, the long-term variations on MHVSR parameters do not seem to respond to the total energy variation on the short-period band of SM.

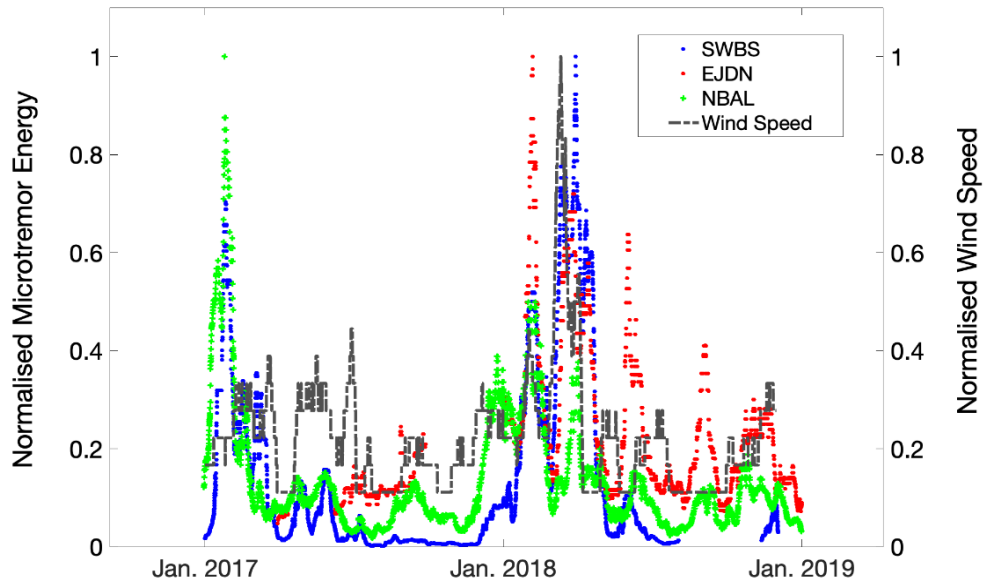

Figure 11. Variation of total microtremor energy in the frequency band [0.2-1.0] Hz for each of the three stations studied in CDB from 2017 to 2019. The wind speed series is over-plotted (black dashed curve).

According to the values in Table 5, the hypothesis of simultaneous SM effects on the MHVSR at the studied stations cannot be rejected when comparing the two studied periods (1) and (2). Whereas in the former period (1) it presents high Pearson correlation coefficients at NBAL and SWBS stations, in the latter it shows lower CC-P values at NBAL and EJDN stations. This could be indicative of microseism mechanisms acting simultaneously in all stations, moving from periods of high to low influence on the MHVSR.

A longer period of time was studied to compare sea-tidal modulations with MHVSR variations. Given that in the moving median filters the window length is the only parameter of choice, the band-pass Butterworth filtering offers wider possibilities for choosing corner periods. Figure 12 exemplifies the differences between these two kinds of filtering processes. The 7-day low-pass filtering shown in the top panels in Figure 12.a does not have a strong difference when compared to the bottom ones done with a 30 days long moving median. However, in Fig. 12.b the Butterworth-type filters demonstrate a better passband performance when looking to restrict to cycles on the scale between 3 and 15 days.

Figure 13 shows the curves of sea tide and MHVSR peak amplitudes compared for the longest window with continuous data available on each seismic station. Up to three bandpass filters were used to find on which temporal scale the SM could be affecting most. It is evidenced how on a prolonged time span the high correlations found during period (1) (Table 5 and Figure 12) disappear. This is in agreement with the work of Becker et al. (2020). These authors found patterns strongly dependent on the season when they compared MHVSR amplitudes and SM. The existence of periods with high correlations and periods with lower ones in Campo de Dalías would be coherent with the SM patterns observed in other coastal locations.

The long-term variations obtained for the MHVSR peaks on Campo de Dalías have been demonstrated to not be in phase with the hypothesised sea tidal action. The low-frequency content of SM would be expected to affect the three seismic stations and their MHVSR peaks equally since there is no evidence of significant structural differences between their locations. Thus, the non-simultaneity of long-term trends from MHVSR variations on the three stations would reject sea tides in Campo de Dalías, and indirectly the SM, as seismic wave sources with a meaningful seasonal influence. However, since the MHVSR peaks found in Campo de Dalías are in a band of likely oceanic action, shorter time intervals disturbed by SM would be justifiably expected.

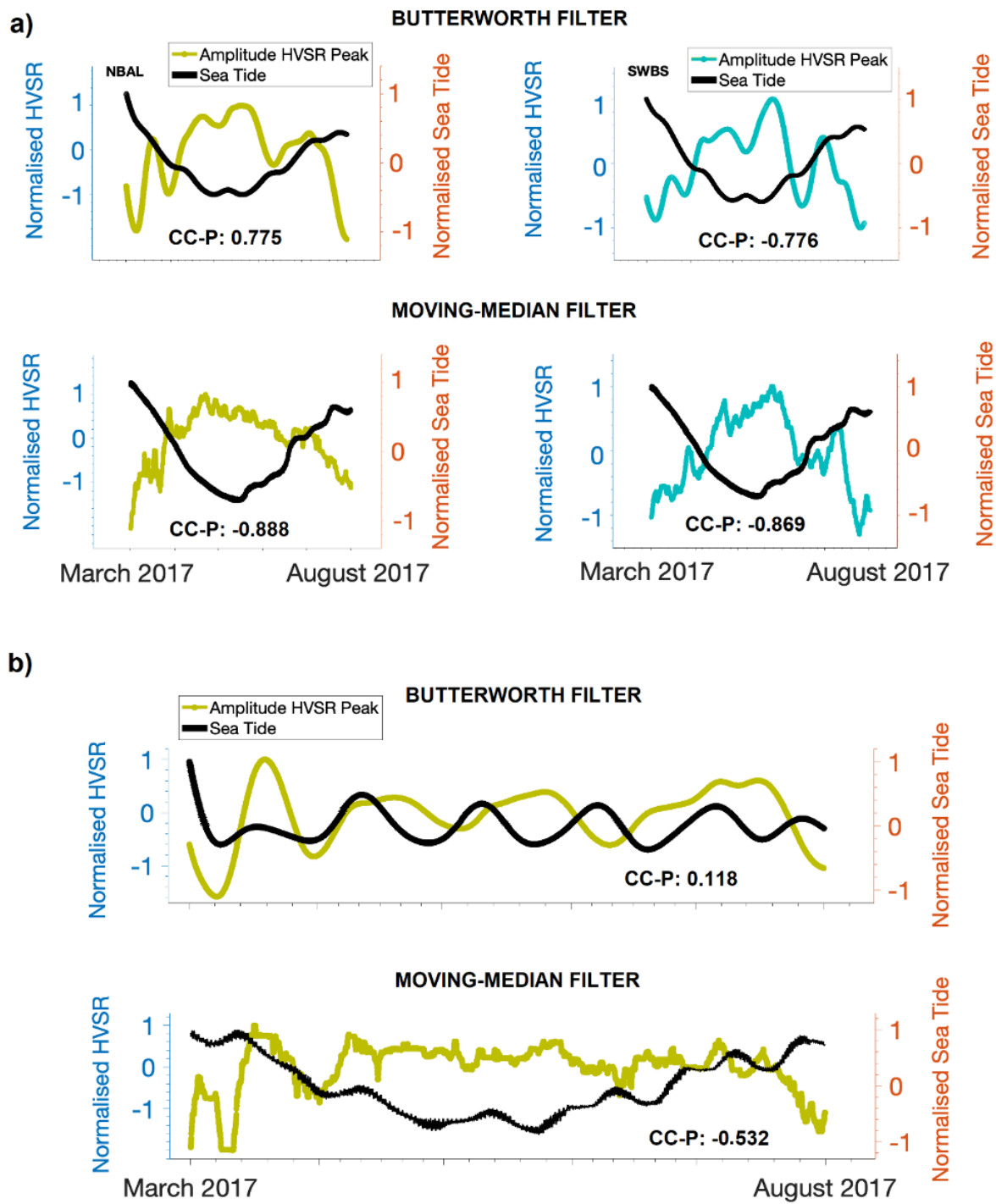

Figure 12. Comparison between the performance of moving median and Butterworth-type filters. a) Top panels show the sea tide and MHVSR amplitude series after applying a second-order low-pass filter with a corner period of 7 days. In the bottom panels, these series were filtered with a moving median filter and sliding window 30 days long. b) The top panel shows the MHVSR amplitude and Sea tide series in NBAL station after being filtered with a second-order band-pass filter with corner periods of 3 and 15 days. The bottom panel shows the same series after applying a moving median filter and sliding window 7 days long. All series were normalised to their maximum value.

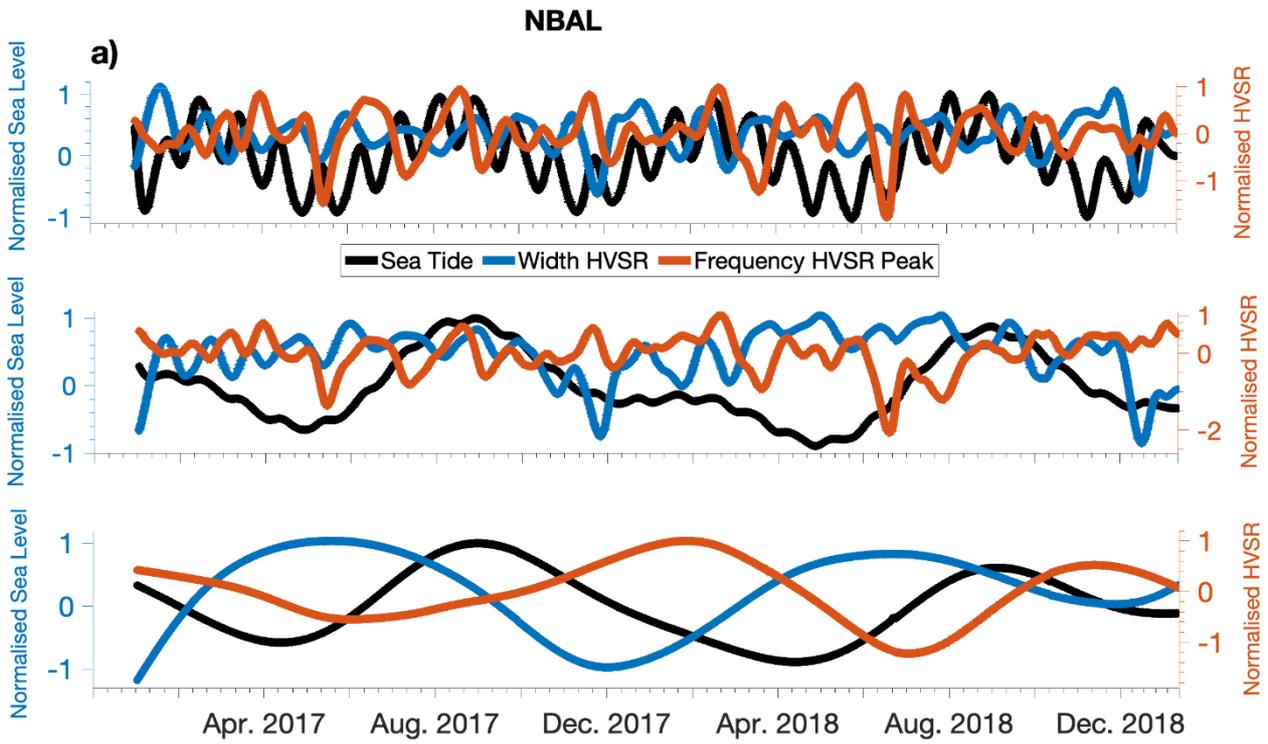

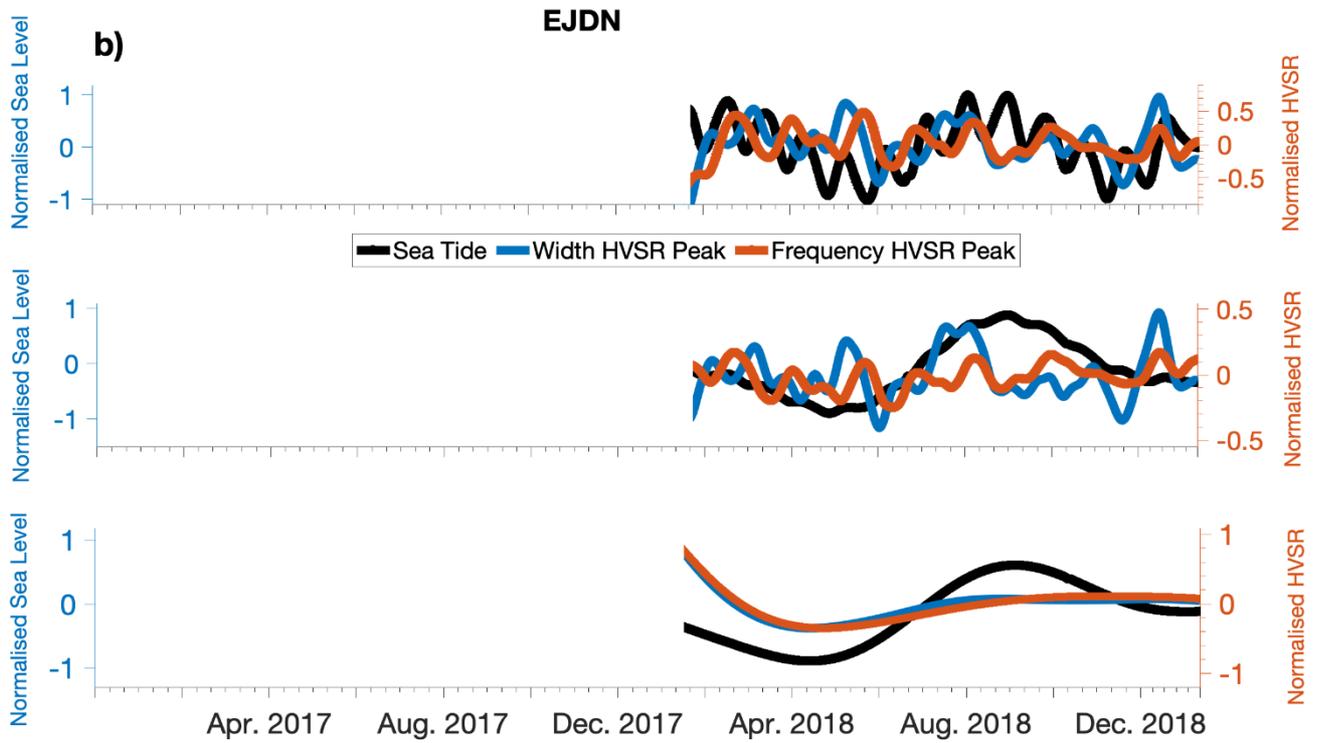

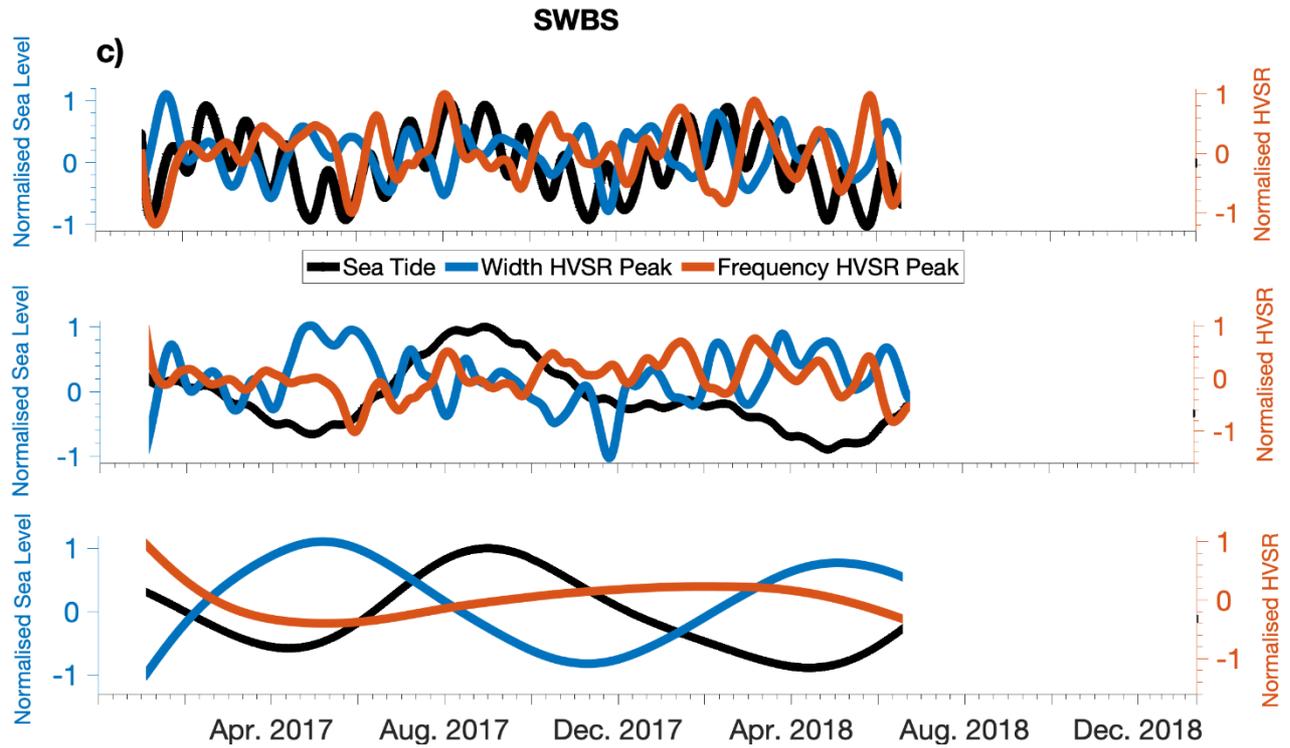

Figure 13. Comparison of sea tide series (black curves) with MHVSR peak-width (blue curves) and peak-amplitude (orange curves) variations after applying three different Butterworth filters for a) NBAL station b) EJDN station and c) SWBS station. The top panels show the results after applying a band-pass filter between 7 and 30 days. Middle panels correspond to a low-pass filter of 7 days. The bottom panels show the results for band-pass filtering between 60 and 180 days. All series are normalised to their maximum value.

Having full continuity of temperature data during 2017 and 2018 led us also to compare temperature seasonality with long-term MHVSR variations. Figure 14 shows the comparison between both variables after using three bandpass filters. Contrary to the sea tide observations, in the narrow band between 7 and 30 days there are several months during which the normalised MHVSR variables closely resemble the normalised temperature. These time intervals are marked with a box in Figure 14, and they are clearly observed in NBAL and SWBS stations. A high positive correlation between the peak frequency and the atmospheric temperature is observed during these windows. The correlation observed for the MHVSR peak-width in these periods is found to be negative instead. Nevertheless, after applying band-pass filtering between 60 and 180 days, the correlation coefficients change their signs for peak-width and peak-frequency in NBAL and SWBS stations. It is observed as well for these two stations that there is a phase shift of two months between the two-time series, temperature and MHVSR variables, filtered between 60 and 180 days.

Thermoelastic responses of the ground have been discussed to be a source mechanism able of varying the seismic velocities or even inducing strain coupled tilts (Hillers et al., 2015; Sthähler et al., 2020). Nevertheless, the observed phase shifts are in contradiction with the characteristic delays between thermoelastic strains and their source temperature field. There is no physical justification for a delay of the temperature with respect to any MHVSR parameter. Thus, in the long term, our data do not show evidence of a causal relationship between the atmospheric temperature and the MHVSR parameters observed in Campo de Dalías.

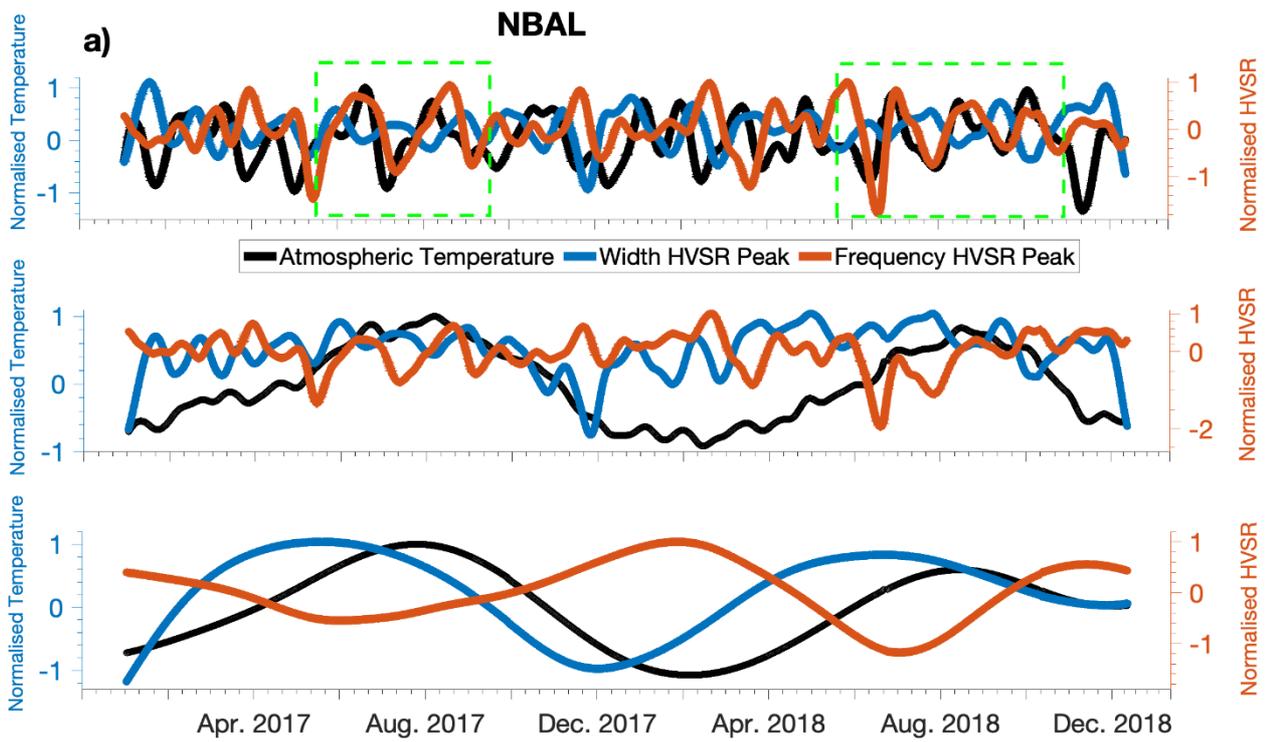

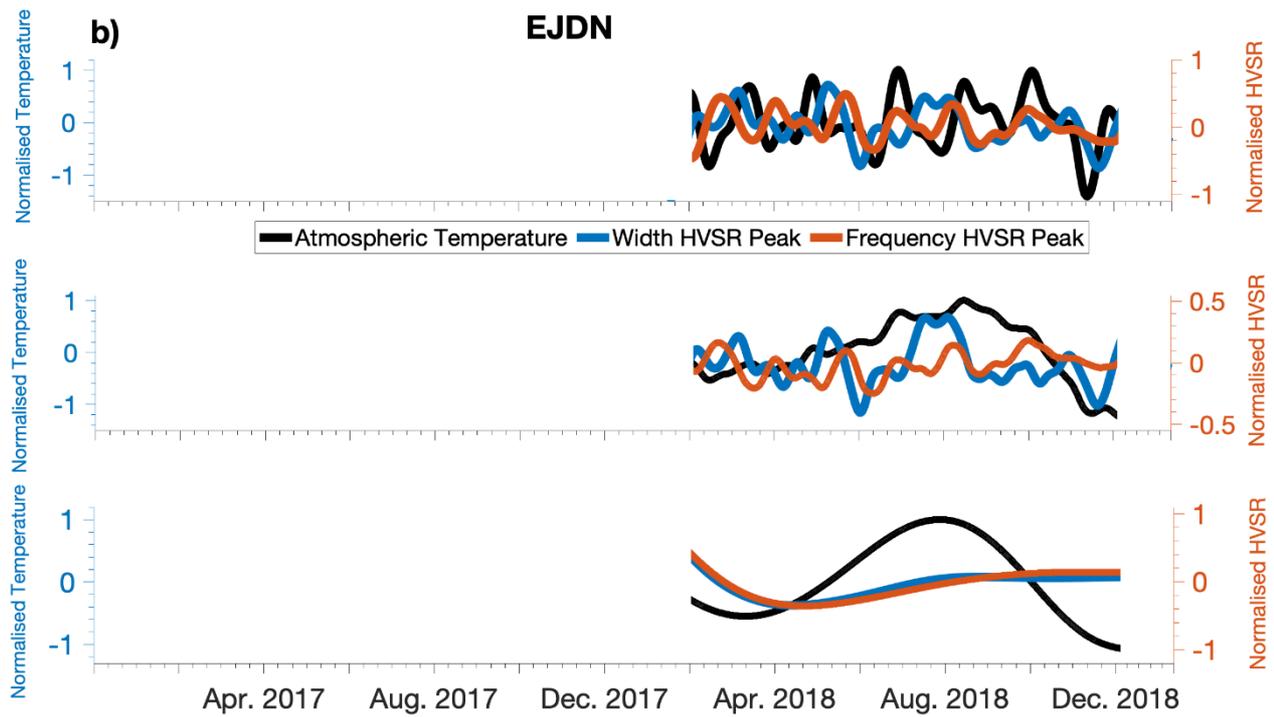

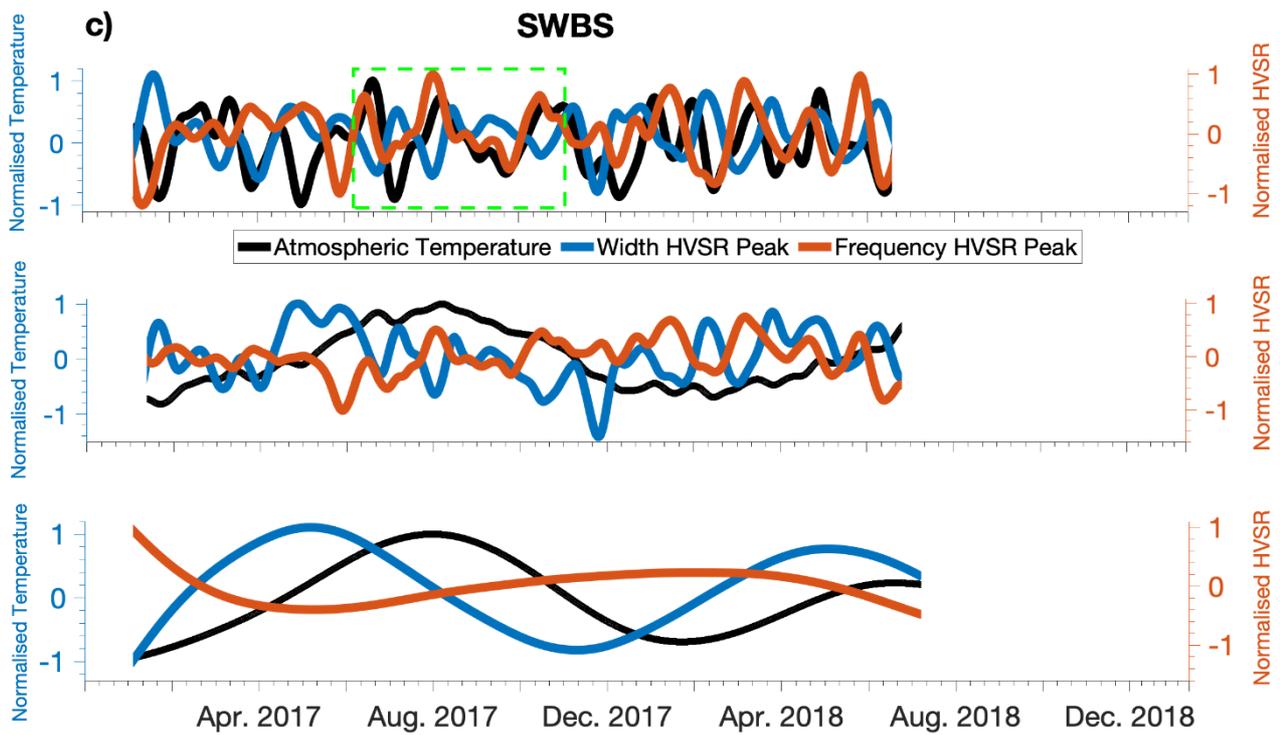

Figure 14. Comparison of atmospheric temperature series (black curves) with MHVSR peak-width (blue curves) and peak-amplitude (orange curves) variations after applying three different Butterworth filters for a) NBAL station b) EJDN station and c) SWBS station. The top panels show the results after applying a band-pass filter between 7 and 30 days. Middle panels correspond to a low-pass filter of 7 days. The bottom panels show the results for band-pass filtering between 60 and 180 days. All series are normalised to their maximum value. Green boxes point to the periods during which the MHVSR peak-frequency variations are in phase and well correlated with temperature variations.

Piezometric data also held significant correlation coefficients with MHVSR parameters (Table 5). But on top of that, and unlike sea tide curves in Figure 12a, the groundwater curves got to share nearly identical trends with some of the MHVSR parameters measured during periods (1) and (2) in Table 5. Such similarities can be seen in Figure 15. The very recent work of Rigo et al. (2021) found that their MHVSR peak amplitudes and resonance frequencies varied seasonally in phase with the aquifer cycles from their study area. The study period by these authors spanned two years. Figure 16 shows the comparison between MHVSR and groundwater series for our longest periods with simultaneous acquisition of piezometric and seismic ambient noise data. Contrary to the atmospheric temperature and sea tide comparisons, and in concordance with the results from Rigo et al. (2021), the seasonality in the long-term MHVSR variations is with minor delays in phase with the groundwater cycles (Figure 15).

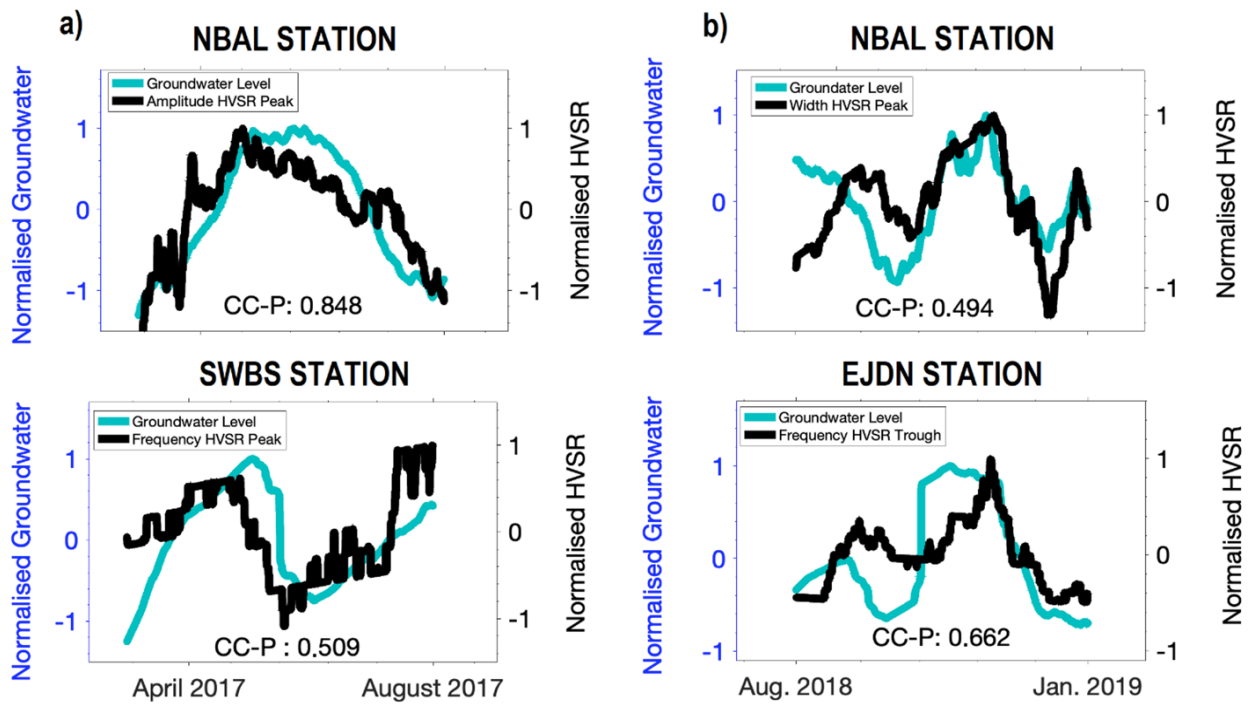

Figure 15. Comparison of groundwater series (black curves) and MHVSR parameters (blue curves) which got the highest CC-P values in Table 5 for periods a) between April and August 2017 and b) between August 2018 and January 2019. These series were normalised to their maximum values after being filtered by a moving median filter with a window size of 30 days.

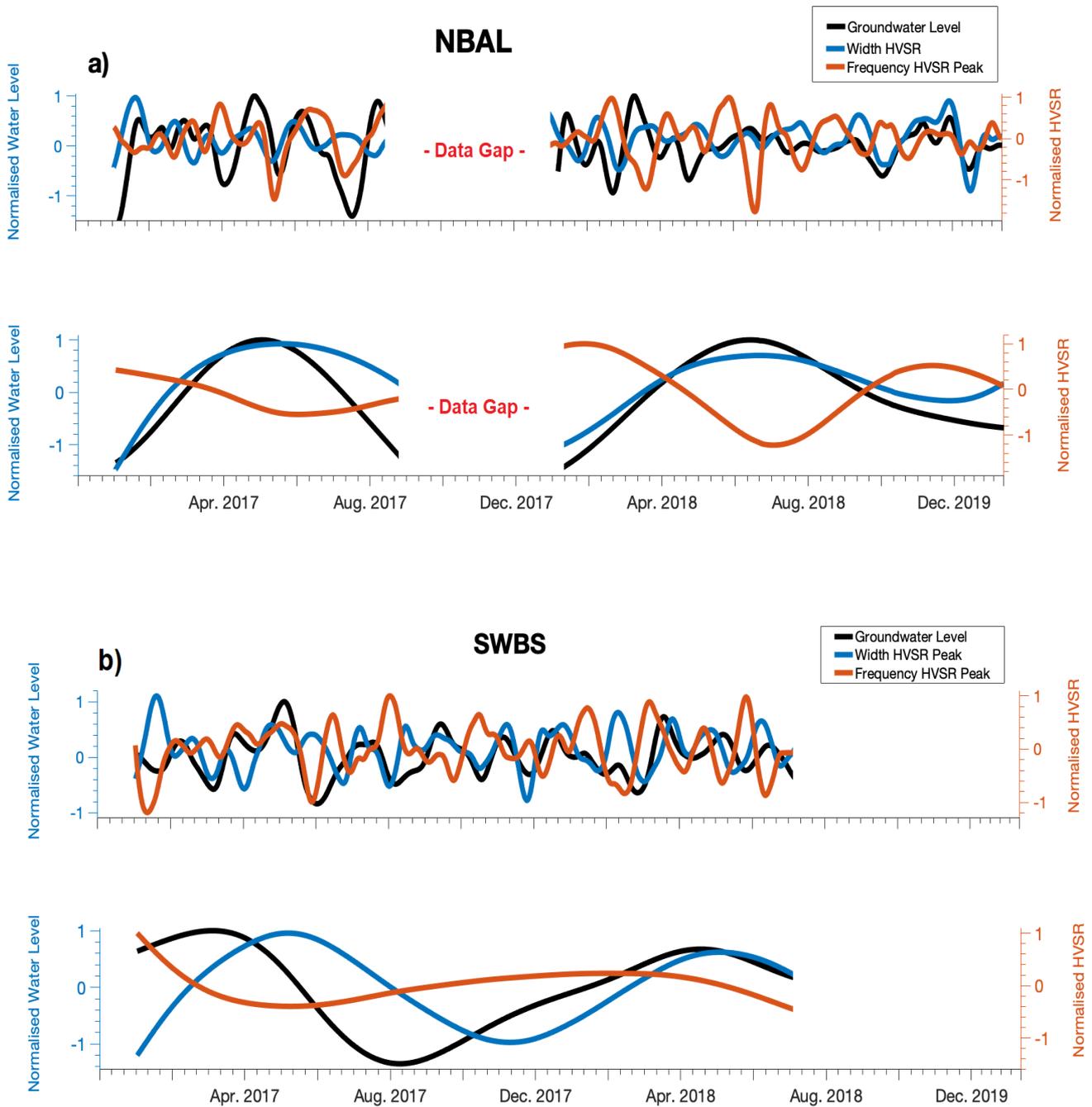

Figure 16. Comparison of groundwater series (black curves) with MHVSR peak-width (blue curves) and peak-amplitude (orange curves) variations after applying two different Butterworth filters for a) NBAL station and b) SWBS station. The top panels show the results after applying a band-pass filter between 7 and 30 days. The bottom panels show the results for band-pass filtering between 60 and 180 days. All series are normalised to their maximum value.

The periodicities, in both MHVSR and groundwater curves, observed after the bandpass filtering between 7 and 30 days are also in phase on NBAL and SWBS stations. Thus, there exists a relationship not only in their long-term seasonality but also in a medium-term time scale. Moreover, the sign of the correlation is kept in both medium and long-term scales. The MHVSR peak-width is kept positively related while the peak-frequency does it negatively after respective bandpass filtering in NBAL and SWBS (Figure 16). Figure 17 a-c shows the MHVSR width- and trough-variations overplotted with piezometric data to facilitate seeing the synchronicities observed in the medium timescale, between 7 and 30 days.

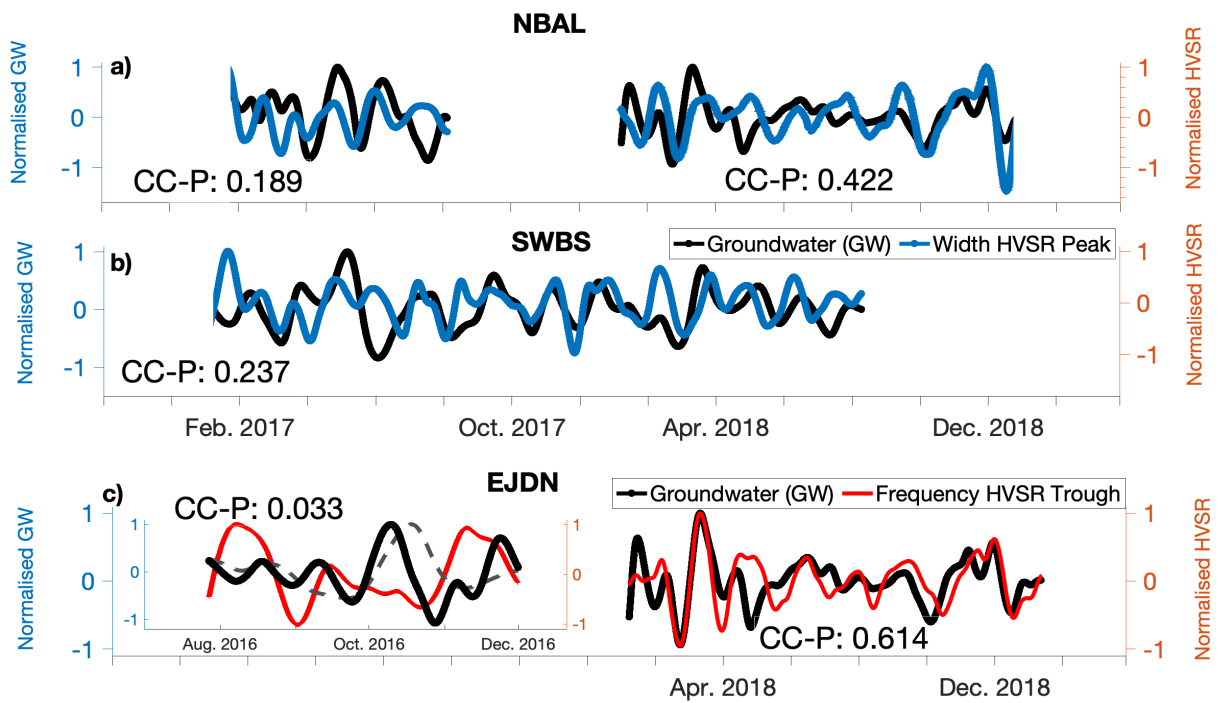

Figure 17. Comparison of groundwater series (black curves) with a) MHVSR peak-width (blue curves) for NBAL station b) MHVSR peak-width (blue curves) for SWBS station and c) frequency of MHVSR trough (red line) for EJDN station. It should be noted that in the comparison of EJDN variations in c) the GW series correspond to the piezometer observations in NBAL (black solid line). The dashed line in c) corresponds to the piezometer observation in EJDN. All series are normalised to their maximum values after having applied a band-pass filter between 7 and 30 days.

Only the groundwater cycle in EJDN showed to have a significant relationship with the MHVSR trough at this station (see Table 5 together with Figures 15.b and 17.c). Piezometric data at that station correspond to the deep aquifer (AQ2 in Fig. 5). Nevertheless, according to Spanish Geological Survey (IGME, 2014) the shallow aquifer AQ1 (which is sampled at stations NBAL and SWBS) should be also present at EJDN. The lack of longer piezometric observations of AQ2 with no missing data impeded us from observing if the MHVSR trough also keeps or not a high correlation with the groundwater variation at EJDN borehole. The comparison between MHVSR trough variations in EJDN with the groundwater level in NBAL was evaluated in two periods: the first one between mid and late 2016 and the second one from early to late 2018 (Figure 17.c). This evaluation proved a high and positive relationship between these two parameters during a one-year period, corresponding to 2018. However, the obtained CC-P for the period between August and December 2016 is as low as 0.03.

There also exist clear periodicities observed in the MHVSR frequency-trough from early 2017 to mid-2018 in the SWBS station that are not synchronised with the other MHVSR variations (Figure 18). The variations observed in the frequency of MHVSR trough in SWBS station did not reveal strong relationships with any of the climate variables studied. The occurrence of the deep aquifer systems in Campo de Dalías is favoured by variable levels of permeability together with fractured conditions of the permeable carbonate rocks (dolostones and limestones) in the Triassic basement. The confinement of these deep aquifers is driven majorly by a series of Pliocene-Miocene marls (see Figure 5). Late-Miocene limestones and calcarenites also host these deep carbonate aquifers, getting to form a connected aquifer with the Triassic basement in those sites where does not exist an intermediate unit of phyllites. Therefore, the deep aquifer would be confined within layers (4) and (5) of our 1-D models (see the position of AQ. 2 in Figure 5).

There exist as well MHVSR trough variations observed in SWBS from early 2017 to mid-2018 that do not correlate with any of the investigated environmental variables (Figure 18). After observing the relationship between the groundwater levels of the shallower aquifer (AQ1 Fig. 5) and the other MHVSR peak parameters in NBAL and EJDN stations, variations from MHVSR trough in SWBS station are an open question for further research. The cycles of discharge of the aquifers in Campo de Dalías are in no way natural due to the heavy pumping rates for agriculture and urban demands. It would be needed to record the piezometric levels to assess if there is some influence from the deep aquifer in the variations experienced by the trough frequency at SWBS station.

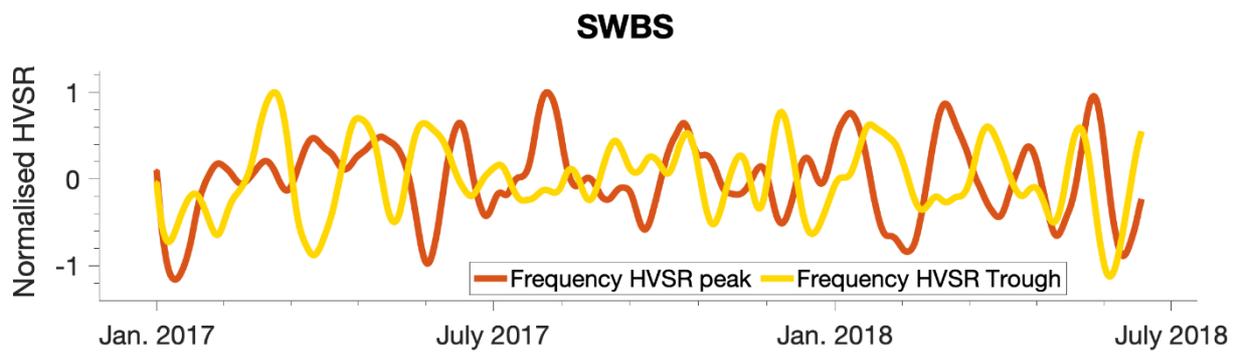

Figure 18. Experimental MHVSR variations in SWBS station observed in the peak-frequency (orange line) and in the trough-frequency (yellow line) after filtering the raw series with a second-order Butterworth bandpass filter between 7 and 30 days. These series are normalised to their maximum values.

3.2.2 Comparison with synthetic observations

The four MHVSR peak parameters did not respond proportionally under variations in the same layer in none of the three stations investigated. Only the MHVSR peak width and peak amplitude appeared to respond proportionally (as a first approximation), both experimentally (Figure 10) and to most of the synthetic model variations (Tables 2 and 3), in the MHVSR clear peak typologies of NBAL and SWBS stations. The broad MHVSR peak observed in EJDN and its less stable behaviour (see Figure 8) causes this MHVSR to be prone to non-proportional variations between its peak amplitude and peak width (Figure 10).

Nonetheless, assuming a multilayer medium only altered by velocity changes as in the synthetic trials in section 3.1, each MHVSR parameter in some cases (see Table 2) might behave independently from the others. Where the changes are occurring is what could hold the key to understanding the individual behaviours from MHVSR parameters. This would provide a reasonable justification of why all the MHVSR parameters do not show identical trends in the experimental data analysed (see e.g., Figures 10 and 18). This sensitivity analysis is exemplified in Figure 19, where some synthetic velocity changes and the behaviour from each MHVSR parameter are shown.

According to the synthetic observations in Figure 19, plus Tables 2 and 3, some conjectures about the origin of our MHVSR variations can be made. The two MHVSR peak parameters, width and amplitude, would be more sensitive to velocity changes occurring in layer (2) for the 1-D model of NBAL (Figure 19a). This happens either in conditions of a constant or a variable Poisson coefficient (Tables 2 and 3). Such behaviour is also observed in the NBAL model when layer (2) is varied simultaneously with other layers. Changing the seismic velocities of layer (2) is the required condition, except when combining layers (3), (4), and (5), to obtain appreciable peak variation to small relative changes.

According to what is observed in Figure 19.a, MHVSR peak-width and peak-amplitude are inversely related to seismic velocity changes occurring in layer (2). In our real data for NBAL these two MHVSR parameters were positively related to the groundwater level variation, which occurs in layer (2). These two behaviours are coherent with the relationship between S-wave velocities and water saturation, by which the former generally decreases when the latter increases (Gassmann, 1951; Baechle et al., 2009).

The MHVSR peak-frequency would be more sensitive to velocity changes in the medium when they happen in layer (3) of NBAL model and in layers (2) and (3) in the SWBS model. In none of the studied cases, under conditions of fixed thicknesses, the MHVSR peak-frequency at NBAL or SWBS would change if it is not accompanied by a change on the $V_S$ parameter (Table 2). In a multi-layered ground model admitting simplification as a single soft layer underlain by the bedrock, the sensitivity of the MHVSR peak-frequency to relative Vs variations in particular layers will be higher for those layers representing greater S-wave travel time (i.e. higher thickness-to-velocity ratios). This would explain why Vs variations in layer (3) affects in a strong way the MHVSR peaks in the synthetic tests performed (Table 2, Figure 6). The synthetic trials varying only the $V_P$ moved the MHVSR peak-position neither in the SWBS model nor in NBAL. The synchronicities observed in the experimental data between MHVSR peak frequencies and temperature oscillations for NBAL and SWBS stations (green boxes in Figure 14) open the possibility to other mechanisms different from the afore-discussed water saturation.

Connectivity and density of fractures in rocks are mechanisms that have been proved to be responsible for body wave velocity changes without the interference of fluid content (Quiroga et al., 2020). As well, according to Snieder et al. (2002), changes derived from temperature effects on the bulk modulus can modify the seismic velocities of the medium in a reversible way. Based on the syn-thetic tests, these synchronicities with the atmospheric temperature would be pointing to mechanisms acting from ten to a hundred metres deep in NBAL and SWBS models. However, such a range of depths is far from the usual range up to which the daily and annual fluctuations of air temperature modify the soil temperature. Therefore, a complex relationship is expected to exist between the air temperature and the MHVSR variability to reach a satisfactory explanation for the observed synchronicities.

The same hypotheses reasoned for NBAL and SWBS relating groundwater level of the shallower aquifer with MHVSR variations apply to the EJDN station. The trough position in EJDN model (Table 1) is conditioned by $V_S$ modifications in synthetic trials, in which it is observed to be more sensitive to layer (3). These synthetic modifications showed that the MHVSR trough of EJDN is positively related to the $V_s$ modifications tried in layer (3) (Figure 19.c). The variation of seismic velocities in all layers of the EJDN model would be necessary to find the MHVSR trough negatively related to them. This variation follows the previous reasoning for NBAL station: the decrease in seismic velocities obeys the increase in water saturation. Variation of seismic velocities in all layers of EJDN model would be possible if the piezometric level of the deeper aquifer (Figure 5) was the main factor controlling the moisture conditions of the subsoil structure.

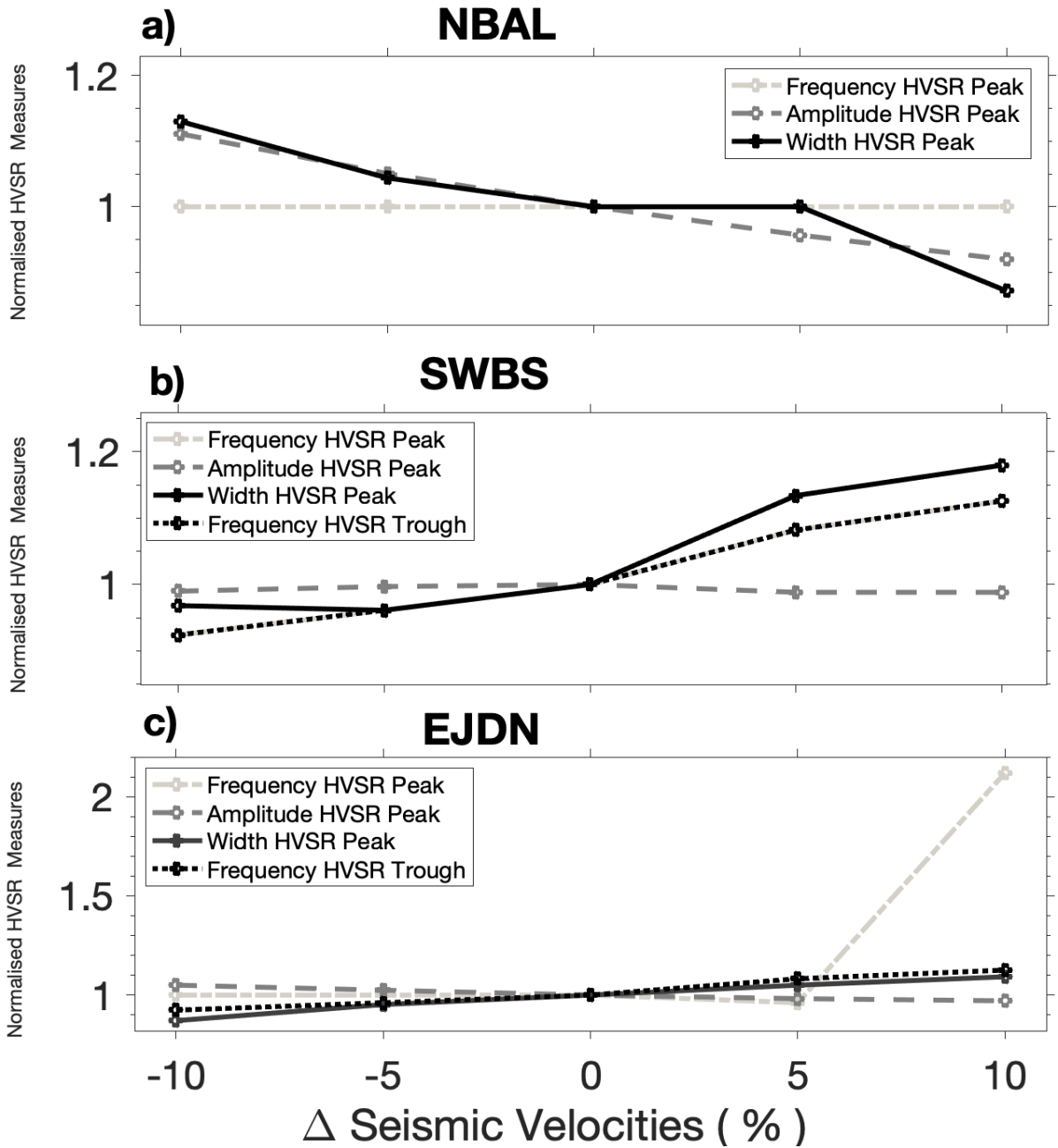

Figure 19. Synthetic variations observed in the MHVSR peaks when varying seismic velocities of: a) layer number (2) in the model of NBAL station; b) all layers in the model of SWBS station and c) layer (3) in the model of EJDN station. The seismic velocities were modified keeping the original Poisson coefficients invariant from models in Table 1. The MHVSR measures are normalised to their unperturbed values.

## 4 Conclusions

Assessment of the seismic site response has encountered in passive methodologies practical and efficient tools with manifold advantages. Since the reliability of their estimations is conditioned under seasonal repeatability, a stationarity analysis can help to confirm or rule out the robustness of results from these methodologies. The results of this study, which combines real data and synthetic sensitivity analyses, suggest that MHVSR peak variations can be a tool for tracking seismic velocity changes in the medium even though such variations follow complicated patterns.

The standard deviations of the fundamental frequency determined by the clear-peak typologies of NBAL and SWBS stations were found to be 8% for both stations in the analysed months (Table 4). The broad character of the MHVSR peak in EJDN station is reflected in a higher percentage of variation, up to 18% for the MHVSR peak frequency. However, these deviations comply with the threshold values recommended by SESAME for reliable identification of the site predominant frequency. Thus, according to the results presented in this paper, MHVSR variations within the margins of reliability may also be wrapping information about seismic velocity changes directly linked to water content and other mechanisms. This adds to the MHVSR, without breaking its robustness for site response analysis, new exploratory capabilities by single station observations.

Traditionally assessed by the variations in peak frequency and amplitude, we incorporated definitions of peak-width and trough frequency as new parameters in the study of MHVSR stability. Taking reference amplitudes to define the MHVSR peak-width complied sufficiently with the purpose of our research to determine width variability for each MHVSR curve analysed (see Fig. 3). On the other hand, the trough variability has been found to be useful also to discern whether or not the shear wave velocities were being altered.

The main conclusions of the tests performed in this article are:

- Among the local climatic and hydrogeological variables and the marine tides for which a linear relationship with the peak parameters of the MHVSR was tested, the water table was the one that maintained it on long and medium-term scales.

-The observed synthetic and real MHVSR shape-behaviours are coherent with the existent relationship between S-wave velocities and water saturation. They are consistent as well with the aquifer position hosted within the sedimentary units of CDB.

- Seasonalities in the long-term MHVSR variations are found to be, with minor delays, in phase with groundwater cycles.

- A significant correlation between the seismic energy and wind speed was found in CDB on the short-period band of SM.

-Long-term variations on MHVSR parameters in CDB are not affected by SM energy. However, in shorter time intervals with SM transients, MHVSR shape-variations would be reasonably expected.

-A straightforward relationship between air-temperature and deep soil mechanisms is not expected to explain the medium-term synchronicities observed between this variable and MHVSR peak-frequency variations. As marked by synthetic tests, the range of depths where velocity variations would be needed to modify the MHVSR shape are far from the usual range of depths where this climate variable is known to induce soil modifications.

It is clear at this point that more than one mechanism is acting on the MHVSR variations observed. An interplay between different mechanisms is expected to be affecting the seismic velocity structure in each study site of CDB. The direct measure of velocity variations would be a necessary incorporation to understand the soil mechanisms and bound their depth extents. Complementing our results with multistation techniques and longer acquisition periods will make a definite contribution to the results found in this research. Longer observations would enable us to see if the observed synchronicities between the MHVSR parameters and weather factors are cyclically repeated on a yearly basis. Besides, the cross-correlation-based techniques would help to quantify the actual seismic velocity rate changes. The MHVSR capabilities would be so positioned as a complementary tool to be easily implemented in geophysical exploration of environments as hydrologically complicated as the karstic aquifers of Campo de Dalías.

Acknowledgements

The piezometric and seismic ambient noise data were provided by the project number CGL2014-59908 funded by Spanish Ministry of Economy and Competitiveness and the European Regional Development Fund. Some results have been obtained from data supplied by Agencia Estatal de Meteorología (government of Spain). The first author thanks her parents for their endless encouragement and rooting for her so unconditionally. Likewise, thanks to João Fontiela, whose insightful comments helped to improve the paper.

Data Availability

The data underlying this article will be shared on reasonable request to the corresponding author.